\documentclass[aps,superscriptaddress,twocolumn,dvips]{revtex4}
\usepackage{feynmp}
\usepackage{amsmath}
\usepackage{amsfonts}
\usepackage{amssymb}
\usepackage{epsfig}
\usepackage{graphicx}
\usepackage{subfigure}
\usepackage{hyperref}
\usepackage{bbm}
\usepackage{color}
\usepackage{longtable}
\usepackage{booktabs}
\usepackage{multirow}
\newcommand{\tabincell}[2]{\begin{tabular}{@{}#1@{}}#2\end{tabular}}

\newcommand{\Rmnum}[1]{\expandafter\@slowromancap\romannumeral #1@}
\allowdisplaybreaks[3]
\begin{document}

\title{Quantum critical phenomena of excitonic insulating transition in two dimensions}

\author{Xiao-Yin Pan}
\affiliation{Department of Physics, Ningbo University, Ningbo,
Zhejiang 315211, China}
\author{Jing-Rong Wang}
\affiliation{Anhui Province Key Laboratory of Condensed Matter
Physics at Extreme Conditions, High Magnetic Field Laboratory of the
Chinese Academy of Sciences, Hefei, Anhui 230031, China}
\author{Guo-Zhu Liu}
\altaffiliation{Corresponding author: gzliu@ustc.edu.cn}
\affiliation{Department of Modern Physics, University of Science and
Technology of China, Hefei, Anhui 230026, China}

\begin{abstract}
We study the quantum criticality of the phase transition between
Dirac semimetal and excitonic insulator in two dimensions. Even
though the system has a semimetallic ground state, there are
observable effects of excitonic pairing at finite temperatures
and/or finite energies, provided that the system is in proximity to
excitonic insulating transition. To determine the quantum critical
behavior, we consider three potentially important interactions,
including the Yukawa coupling between Dirac fermions and excitonic
order parameter fluctuation, the long-range Coulomb interaction, and
the disorder scattering. We employ the renormalization group
technique to study how these interactions affect quantum criticality
and also how they influence each other. We first investigate the
Yukawa coupling in the clean limit, and show that it gives rise to
typical non-Fermi liquid behavior. Adding random scalar potential to
the system always turns such a non-Fermi liquid into a compressible
diffusive metal. In comparison, the non-Fermi liquid behavior is
further enhanced by random vector potential, but is nearly
unaffected by random mass. Incorporating the Coulomb interaction may
change the results qualitatively. In particular, the non-Fermi
liquid state is protected by the Coulomb interaction for weak random
scalar potential, and it becomes a diffusive metal only when random
scalar potential becomes sufficiently strong. When random vector
potential or random mass coexists with Yukawa coupling and Coulomb
interaction, the system is stable non-Fermi liquid state, with
fermion velocities flowing to constants in the former case and being
singularly renormalized in the latter case. These quantum critical
phenomena can be probed by measuring observable quantities. We also
find that, while the fermion velocity anisotropy is not altered by
the excitonic quantum fluctuation, it may be driven by the Coulomb
interaction to flow to the isotropic limit.
\end{abstract}

\maketitle


\section{Introduction}

In the past decade, the unconventional properties of various
Dirac/Weyl semimetal (SM) materials \cite{Sarma11, Kotov12, Vafek14,
Wehling14, Wan11, Weng16, FangChen16, Yan17, Hasan17, Armitage18}
have been investigated extensively. Many of the unconventional
properties are related to the existence of isolated Dirac/Weyl
points, at which the conduction and valence bands touch. When the
chemical potential is tuned to exactly the Dirac points, the fermion
density of states (DOS) vanishes at the Fermi level. As a result,
the Coulomb interaction is long-ranged due to the absence of static
screening. Extensive previous studies \cite{Kotov12, Gonzalez99,
Hofmann14, Sharma16, Goswami11, Hosur12, Hofmann15, Throckmorton15,
Sharma18, Yang14, Abrikosov72, Abrikosov71, Abrikosov74, Moon13,
Herbut14, Janssen15, Dumitrescu15, Janssen17, Huh16, Cho16,
Isobe16B, Lai15, Jian15, Zhang17, WangLiuZhang17MWSM} have revealed
that the Coulomb interaction leads to a variety of unconventional
low-energy behaviors.

Among all the known SM materials, two-dimensional Dirac SM,
abbreviated as 2D DSM hereafter, has been studied most extensively,
usually in the context of graphene. Renormalization group (RG)
analysis \cite{Shankar94, Kotov12} has revealed that the long-range
Coulomb interaction is marginally irrelevant in the weak-coupling
regime. When the Coulomb interaction is strong enough, the
originally massless fermions can acquire a dynamical mass gap via
the formation of stable particle-hole pairs
\cite{CastroNetoPhysics09, Khveshchenko01, Gorbar02, Khveshchenko04,
Liu09, Khveshchenko09, Gamayun10, Sabio10, Zhang11, Liu11,
WangLiu11A, WangLiu11B, WangLiu12, Popovici13, WangLiu14,
Gonzalez15, Carrington16, Sharma17, Xiao17, Carrington18, Gamayun09,
WangJianhui11, Katanin16, Gonzalez10, Gonzalez12, Drut09A, Drut09B,
Drut09C, Armour10, Armour11, Buividovich12, Ulybyshev13, Smith14,
Juan12, Kotikov16, Gonzalez14, Braguta16, Xiao18, Janssen16,
WangLiuZhangSemiDSM}. This gap generating scenario is
non-perturbative, and has the same picture as excitonic pairing, a
notion proposed decades ago \cite{Keldysh64, Jerome67}. In the
special case of 2D DSM, such an excitonic gap dynamically breaks a
continuous chiral (sublattice) symmetry, which can be regarded as a
condensed-matter realization of the dynamical chiral symmetry
breaking \cite{Nambu61, Miransky94}. The finite gap opened at the
Dirac point drives the SM to undergo a quantum phase transition
(QPT) into an excitonic insulator (EI). The EI is induced only when
the effective interaction strength, denoted by $\alpha$, exceeds
some critical value $\alpha_c$, which defines SM-EI quantum critical
point (QCP).

In recent years, the possibility of SM-EI transition in graphene has
been investigated by means of various analytical and numerical
techniques. Early calculations \cite{CastroNetoPhysics09,
Khveshchenko01, Gorbar02, Khveshchenko04, Liu09, Khveshchenko09,
Gamayun10, Drut09A, Drut09B, Drut09C} predicted that the Coulomb
interaction in suspended graphene is strong enough to open an
excitonic gap at zero temperature. Specifically, the critical value
$\alpha_c$ was claimed to be smaller than the physical value $\alpha
= 2.16$. However, no visible experimental evidence for excitonic gap
has been observed at low temperatures \cite{Elias11, Mayorov12}.
More careful numerical calculations \cite{WangLiu12, Gonzalez15,
Carrington16, Ulybyshev13, Smith14, Kotikov16} revealed that the
critical value $\alpha_c$ is actually larger than $2.16$, which
implies that the Coulomb interaction cannot generate a finite
excitonic gap. Owing to the conceptual importance and also the
potential technical applications, theorists are still searching for
possible approaches to promote excitonic pairing in various SM
materials. For instance, it was proposed that excitonic pairing may
be promoted by an additional short-range repulsive interaction
\cite{Liu09, Gamayun10, WangLiu12} or by certain extrinsic effects,
such as strain \cite{Tang15}.

\begin{figure}[htbp]
\center
\includegraphics[width=2.88in]{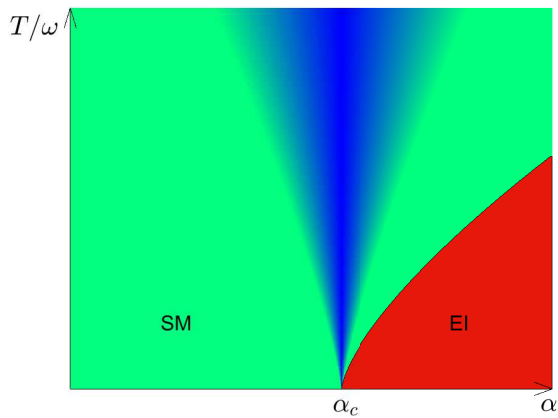}
\caption{Global phase diagram of 2D DSM on the $\alpha$-$T$ or
$\alpha$-$\omega$ plane. Here, $\omega$ stands for the fermion
energy. Deep in the insulating phase, the fermions are suppressed at
low energies. Deep in the semimetallic phase, the Coulomb
interaction is too weak to form excitonic pairs. The excitonic
insulating transition occurs as $\alpha$ increases up to $\alpha_c$
at $T=0$. This point is broadened into a finite quantum critical
regime at finite $T$ and/or finite $\omega$. The excitonic quantum
fluctuation has observable effects in the whole quantum critical
regime.} \label{Fig:PhaseEIQCP}
\end{figure}

Most previous works on SM-EI QPT have focused on the precise
calculation of $\alpha_c$ at zero temperature ($T=0$) by means of
various techniques \cite{CastroNetoPhysics09, Khveshchenko01,
Gorbar02, Khveshchenko04, Liu09, Khveshchenko09, Gamayun10, Sabio10,
Zhang11, Liu11, WangLiu11A, WangLiu11B, WangLiu12, Popovici13,
WangLiu14, Gonzalez15, Carrington16, Sharma17, Xiao17, Carrington18,
Gamayun09, WangJianhui11, Katanin16, Gonzalez10, Gonzalez12,
Drut09A, Drut09B, Drut09C, Armour10, Armour11, Buividovich12,
Ulybyshev13, Smith14, Juan12, Kotikov16, Gonzalez14, Braguta16,
Xiao18, Janssen16, WangLiuZhangSemiDSM}. In this paper, we propose
to explore the signatures of excitonic pairing at finite $T$ and/or
finite energy $\omega$. Here is our logic: even though the exact
zero-$T$ ground state of suspended graphene (or other 2D DSMs) is
gapless, the quantum fluctuation of excitonic pairs still have
observable effects at finite $T$ and/or $\omega$ if the system is in
the quantum critical regime around the putative SM-EI QCP. Recent
Monte Carlo simulations \cite{Ulybyshev13} and Dyson-Schwinger
equation study \cite{Carrington18} both suggest that the value
$\alpha_c$ is not far from the physical value of suspended graphene.
As illustrated in the schematic phase diagram
Fig.~\ref{Fig:PhaseEIQCP}, if $\alpha$ is slightly smaller than
$\alpha_c$, no excitonic gap is opened at $T=0$ and the excitonic
order parameter has a vanishing mean-value. However, the quantum
fluctuation of excitonic order parameter is not negligible at finite
$T$ and/or $\omega$ and may lead to considerable corrections to
observable quantities. For instance, the nuclear-magnetic-resonance
measurements performed by Hirata \emph{et al.} \cite{Hirata17}
indicate that the compound $\alpha$-(BEDT-TTF)$_{2}$I$_{3}$ is close
to an SM-EI QCP and that the excitonic fluctuation results in
singular corrections to the nuclear magnetic resonance relaxation
rate.

We study the quantum critical phenomena emerging in the broad
quantum critical regime around SM-EI QCP, with the aim to explore
observable effects of excitonic pairing. For this purpose, we take
suspended graphene (typical 2D DSM) as our starting model, and
calculate the interaction corrections to some observable quantities
of Dirac fermions. In this regime, the gapless fermions interact
with the quantum critical fluctuation of excitonic order parameter,
which is described by a Yukawa coupling term. The long-range Coulomb
interaction is still present and needs to be properly taken into
account. Moreover, there is always certain amount of quenched
disorder \cite{Kotov12} in realistic materials, and the
fermion-disorder coupling might play a vital role. The actual
quantum critical phenomena cannot be accurately determined if one or
more of these interactions are naively ignored or improperly
treated. We emphasize that, these three kinds of interaction may
have very complicated mutual influence. To make a generic analysis,
we will treat all the three kinds of interaction on equal footing
and study their interplay carefully.

As the first step, we treat the Yukawa coupling in the clean limit,
and demonstrate that this coupling leads to strong violation of
Fermi liquid (FL) theory. Indeed, the quasiparticle residue $Z_f$
vanishes at low energies, and the fermion DOS $\rho(\omega)$
receives power-law corrections from the excitonic fluctuation. Both
of these two features are typical non-Fermi-liquid (NFL) behaviors.
If the fermion dispersion is originally anisotropic, the ratio
between two fermion velocities is unrenormalized.

The next step is to incorporate quenched disorder and analyze its
interplay with the Yukawa coupling. We find that the resultant
low-energy properties depend sensitively on the nature of the
disorder. Adding random scalar potential (RSP) to the system always
turns the NFL caused by Yukawa coupling in the clean limit into a
compressible diffusive metal (CDM). The CDM state is characterized
by the generation of a finite zero-energy fermion DOS and a constant
zero-$T$ disorder scattering rate. Different from RSP, random vector
potential (RVP) tends to further enhance the NFL behavior, whereas
random mass (RM) has negligible effects on the system.

We finally incorporate the Coulomb interaction, and find that it
changes the above results qualitatively. In the case of weak RSP,
the Coulomb interaction suppresses disorder scattering and as such
renders the stability of the NFL state caused by Yukawa coupling.
However, such a NFL is converted into CDM once RSP becomes
sufficiently strong. The combination of Yukawa coupling, Coulomb
interaction, and RVP produces a stable NFL state in which the two
fermion velocities flow to constant values in the zero energy limit.
When the Yukawa coupling, Coulomb interaction, and RM are considered
simultaneously, we show that the Coulomb interaction is marginally
irrelevant and RM is irrelevant. These results indicate that the
true quantum critical phenomena are determined by a delicate
interplay of excitonic fluctuation, Coulomb interaction, and
disorder scattering.

Our results might be applied to understand some 2D DSM materials,
such as uniaxially strained graphene \cite{Sharma17, Xiao17} and
organic compound $\alpha$-(BEDT-TTF)$_{2}$I$_{3}$ \cite{Hirata17}.
In these systems, the fermion velocities along different directions
may be unequal. It is thus necessary to examine how interactions
change the anisotropy. According to our RG analysis, the fermion
velocity anisotropy is unaffected by the excitonic fluctuation, but
could be significantly suppressed by the Coulomb interaction.

The rest of the paper will be arranged as following. The model is
presented in Sec.~\ref{Sec:Model}. The RG equations for the
corresponding parameters are shown in Sec.~\ref{Sec:RGEquations}.
The numerical results for different conditions are given and
analyzed in Sec.~\ref{Sec:NumResults}. The mains results are
summarized in Sec.~\ref{Sec:SummaryDiscuss}. The detailed derivation
of the RG equations can be found in Appendices.

\section{The model\label{Sec:Model}}

The fermion energy dispersion in intrinsic graphene is isotropic. It
becomes anisotropic when graphene is deformed. Generically, the
action of free 2D Dirac fermions with anisotropic dispersion is
given by
\begin{eqnarray}
S_{f} = \sum_{\sigma=1}^{N}\int_{\tau,\mathbf{x}}
\bar{\Psi}_{\sigma}(\tau,\mathbf{x})\left[\partial_{\tau}\gamma_{0}
+ \mathcal{H}_{f}\right] \Psi_{\sigma}(\tau,\mathbf{x}),
\end{eqnarray}
where $\int_{\tau,\mathbf{x}}\equiv\int d\tau\int d^2\mathbf{x}$ and
$\mathcal{H}_{f} = -iv_{1}\nabla_{1}\gamma_{1} -iv_{2}\nabla_{2}
\gamma_{2}$. Here, $\Psi$ is a four-component spinor, and
$\bar{\Psi}=\Psi^{\dag}\gamma_{0}$. The matrices $\gamma_{0,1,2}$
are defined as $\gamma_{0,1,2} =
\left(\tau_{3},-i\tau_{2},i\tau_{1}\right)\otimes\tau_{3}$ in terms
of Pauli matrices $\tau_{i}$ with $i = 1, 2, 3$. The gamma matrices
satisfy the anti-commutative rule
$\left\{\gamma_{\mu},\gamma_{\nu}\right\} =
2\mathrm{diag}(1,-1,-1)$. The fermion species is denoted by
$\sigma$, which sums from $1$ to $N$. Fermion flavor $N$ is assumed
to be a general large integer. We use $v_{1}$ and $v_{2}$ to
represent the fermion velocities along two orthogonal directions.

The action of the quantum fluctuation of excitonic order parameter
can be written as
\begin{eqnarray}
S_{b} = \int_{\tau,\mathbf{x}}\left[\frac{1}{2}
\left(\partial_{\tau}\phi\right)^{2}+\frac{c^{2}}{2}
\left(\mathbf{\nabla}\phi\right)^{2} + \frac{r}{2}\phi^{2} +
\frac{u}{24}\phi^{4}\right],
\end{eqnarray}
where $c$ is the boson velocity. Varying boson mass $r$ tunes the
QPT between SM and EI phases. At the QCP, the mass vanishes, i.e.,
$r=0$, and the boson field $\phi$ describes the quantum critical
fluctuation of excitonic order parameter. The quartic
self-interacting term has a coupling constant $u$. The Yukakwa
coupling between fermions and excitonic order parameter is given by
\begin{eqnarray}
S_{fb} = \lambda\sum_{\sigma=1}^{N}\int_{\tau,\mathbf{x}}
\phi\bar{\Psi}_{\sigma}\Psi_{\sigma},
\end{eqnarray}
where $\lambda$ is the corresponding coupling constant.

The excitonic pairing originates from the Coulomb interaction
between fermions and their anti-fermions (holes). Inside the EI
phase, a finite gap is opened at the Fermi level and strongly
suppresses the low-energy fermion DOS. In this case, the Coulomb
interaction and even the fermionic degrees of freedom can be
neglected, and the low-energy properties of the EI phase is mainly
governed by the dynamics of neutral excitons. In contrast, the
fermions remain gapless at the SM-EI QCP. The Coulomb interaction
between gapless fermions may play an important role at low energies.
The action for Coulomb interaction is described by
\begin{eqnarray}
S_{ee}=\frac{1}{4\pi}\sum_{\sigma,\sigma'=1}^{N}
\int_{\tau,\mathbf{x},\mathbf{x}'}\rho_{\sigma}(\tau,\mathbf{x})
\frac{e^2/\epsilon}{|\mathbf{x}-\mathbf{x}'|}
\rho_{\sigma}(\tau,\mathbf{x}'),
\end{eqnarray}
where $\int_{\tau,\mathbf{x},\mathbf{x}'}\equiv\int d\tau\int
d^2\mathbf{x}\int d^2\mathbf{x}'$. The fermion density operator is
defined as $\rho_{\sigma}(\tau,\mathbf{x}) =
\bar{\Psi}_{\sigma}(\tau,\mathbf{x})\gamma_{0}
\Psi_{\sigma}(\tau,\mathbf{x})$. In addition, $e$ is electric charge
and $\epsilon$ dielectric constant.

Disorder exists in almost all realistic materials. Many of the
low-energy behaviors of fermions are heavily affected by disorder
scattering, especially at low $T$. The fermion-disorder coupling is
formally described by
\begin{eqnarray}
S_{\mathrm{dis}} = v_{\Gamma}\int d\tau d^2\mathbf{x}
\bar{\Psi}_{\sigma}(\mathbf{x})\Gamma\Psi_{\sigma}(\mathbf{x})
A(\mathbf{x}).
\end{eqnarray}
The random field $A(\mathbf{x})$ is assumed to be a Gaussian white
noise, i.e., $\langle A(\mathbf{x})\rangle = 0$ and $\langle
A(\mathbf{x})A(\mathbf{x}')\rangle = \Delta
\delta^{2}(\mathbf{x}-\mathbf{x}')$. Here, $\Delta$ is the impurity
concentration, and $v_{\Gamma}$ measures the strength of a single
impurity. The disorders are classified by the expression of $\Gamma$
matrix \cite{Ludwig94, Nersesyan95, Altland02}. For
$\Gamma_{0}=\gamma_{0}$, $A(\mathbf{x})$ is a RSP. For
$\Gamma_{j}=\mathbbm{1}_{4\time4}$, $A(\mathbf{x})$ serves as a RM.
In comparison, RVP has two components $A_{1,2}(\mathbf{x})$,
characterized by $\Gamma=(\gamma_{1},\gamma_{2})$ and
$v_{\Gamma}=(v_{\Gamma1}, v_{\Gamma2})$.

The free fermion propagator has the form
\begin{eqnarray}
G_{0}(\omega,\mathbf{k}) = \frac{1}{-i\omega\gamma_{0} +
v_{1}k_{1}\gamma_{1}+v_{1}k_{2}\gamma_{2}}.
\label{Eq:FermionPropagator}
\end{eqnarray}
The Yukawa coupling can be treated by the RG method in combination
with the $1/N$ expansion. Following the scheme developed by Huh and
Sachdev \cite{Huh08}, we re-scale $\phi$ and $r$ as follows:
$\phi\rightarrow\phi/\lambda$ and $r\rightarrow Nr\lambda^{2}$.
Accordingly, the bare propagator of $\phi$ is expressed as
\begin{eqnarray}
D_{0}^{A}(\Omega,\mathbf{q})=\frac{1}{\frac{\Omega^{2} +
c^2\mathbf{q}^{2}}{\lambda^{2}}+Nr}.
\end{eqnarray}
Near the QCP, we take $r=0$ and then get
\begin{eqnarray}
D_{0}^{A}(\Omega,\mathbf{q}) = \frac{\lambda^{2}}{\Omega^{2} +
c^2\mathbf{q}^{2}}.
\end{eqnarray}
The free boson propagator is drastically altered by the polarization
function, which, to the leading order of $1/N$ expansion, is
\begin{eqnarray}
\Pi^{A}(\Omega,\mathbf{q}) &=& N \int\frac{d\omega}{2\pi}\frac{d^2
\mathbf{k}}{(2\pi)^2} \nonumber \\
&& \times \mathrm{Tr} \left[G_{0}(\omega,\mathbf{k})
G_{0}(\omega+\Omega,\mathbf{k}+\mathbf{q})\right]
\nonumber \\
&=& \frac{N}{4v_1 v_2}\sqrt{\Omega^2 + v_{1}^{2}q_{1}^{2} +
v_{2}^{2} q_{2}^{2}}.
\end{eqnarray}
Now the dressed boson propagator becomes
\begin{eqnarray}
D^{A}(\Omega,\mathbf{q})=\frac{1}{\frac{\Omega^{2} + c^2
\mathbf{q}^{2}}{\lambda^{2}}+\Pi^{A}(\Omega,\mathbf{q})}.
\end{eqnarray}
It is obvious that $\Pi^{A}$ dominates over the free term in the
low-energy regime. Thus, the above expression can be further
simplified to
\begin{eqnarray}
D^{A}(\Omega,\mathbf{q}) \approx \frac{1}{\Pi^{A}(\Omega,\mathbf{q})}.
\end{eqnarray}

The bare Coulomb interaction is described by
\begin{eqnarray}
D_{0}^{B}(\mathbf{q})=\frac{2\pi e^{2}}{\epsilon|\mathbf{q}|}.
\end{eqnarray}
The dynamical screening is encoded in the polarization
$\Pi^{B}(\Omega,\mathbf{q})$, whose leading order expression is given by
\begin{eqnarray}
\Pi^{B}(\Omega,\mathbf{q}) &=& -N \int \frac{d\omega}{2\pi}
\frac{d^2 \mathbf{k}}{(2\pi)^2} \mathrm{Tr}
\left[\gamma_{0}G_{0}(\omega,\mathbf{k})\gamma_{0}\right.\nonumber
\\
&&\left.\times G_{0}(\omega+\Omega,\mathbf{k}+\mathbf{q})\right]
\nonumber \\
&=& \frac{N}{8v_1v_2}\frac{v_{1}^2q_{1}^{2} + v_{2}^2
q_{2}^{2}}{\sqrt{\Omega^2+v_{1}^{2}q_{1}^{2}+v_{2}^{2}q_{2}^{2}}}.
\end{eqnarray}
The dressed Coulomb interaction can be written as
\begin{eqnarray}
D^{B}(\Omega,\mathbf{q}) = \frac{1}{\frac{\epsilon|\mathbf{q}|}{2\pi
e^{2}} + \Pi^{B}(\Omega,\mathbf{q})}.
\end{eqnarray}

In previous works on the quantum criticality of SM-EI transition,
the interplay of Yukawa coupling, Coulomb interaction, and disorder
has never been systematically studied. Here, we emphasize that all
the three interactions could be very important at low energies and
thus should be treated equally.

\section{Renormalization group equations \label{Sec:RGEquations}}

The interplay of distinct interactions can be handled by means of
perturbative RG approach. The detailed RG calculations are presented
in the Appendices. In this section, we only list the coupled RG
equations of a number of model parameters and then analyze their
low-energy properties. The effective model contains several
independent parameters, such as $v_1$, $v_2$, and $v_\Gamma$. These
parameters are renormalized by interactions. To specify how the
interactions alter the fermion dispersion anisotropy, we need to
determine the flow of the ratio $v_2/v_1$. Moreover, to judge
whether FL theory is applicable, we should compute the flow equation
of the residue $Z_f$.

After incorporating three types of interaction in a self-consistent
way, we find that the coupled RG equations for $Z_f$, $v_{1}$,
$v_{2}$, and $v_{2}/v_{1}$ are given by
\begin{eqnarray}
\frac{dZ_{f}}{d\ell} &=& \left(C_{0}^{A}+C_{0}^{B} -
C_{g}\right)Z_{f}, \label{Eq:VRGZf} \\
\frac{dv_{1}}{d\ell} &=& \left(C_{0}^{A} + C_{0}^{B} - C_{1}^{A} -
C_{1}^{B} - C_{g}\right)v_{1}, \label{Eq:VRGV1} \\
\frac{dv_{2}}{d\ell} &=& \left(C_{0}^{A}+C_{0}^{B} - C_{2}^{A} -
C_{2}^{B} - C_{g}\right)v_{2}, \label{Eq:VRGV2}
\\
\frac{d\left(v_{2}/v_{1}\right)}{d\ell} &=& \left(C_{1}^{A} -
C_{2}^{A} + C_{1}^{B} - C_{2}^{B}\right)\frac{v_{2}}{v_{1}}.
\label{Eq:VRGVRatio}
\end{eqnarray}
RG analysis is performed by integrating out the modes defined within
the momentum shell $e^{-\ell}\Lambda < |\mathbf{k}| < \Lambda$,
where $\Lambda$ is an UV cutoff and $\ell$ is a running parameter
\cite{Shankar94}. The lowest energy limit is reached as $\ell
\rightarrow \infty$. For RSP, the flow equation of $v_{\Gamma}$
takes the form
\begin{eqnarray}
\frac{d v_{\Gamma}}{d\ell}=0. \label{Eq:VRGVGammaRCP}
\end{eqnarray}
For the two components of RVP, the flow equations for $v_{\Gamma1}$
and $v_{\Gamma2}$ are
\begin{eqnarray}
\frac{dv_{\Gamma1}}{d\ell}=\left(C_{0}^{A}+C_{0}^{B}-C_{1}^{A}
-C_{1}^{B}-C_{g}\right)v_{\Gamma1}, \label{Eq:VRGVGammaRVP1}
\\
\frac{dv_{\Gamma2}}{d\ell}=\left(C_{0}^{A}+C_{0}^{B}-C_{2}^{A}
-C_{2}^{B}-C_{g}\right)v_{\Gamma2}. \label{Eq:VRGVGammaRVP2}
\end{eqnarray}
For RM, the flow equation of $v_{\Gamma}$ is
\begin{eqnarray}
\frac{dv_{\Gamma}}{d\ell}&=&\left(2C_{0}^{A}+C_{1}^{A}+C_{2}^{A} +
2C_{0}^{B}-C_{1}^{B}-C_{2}^{B}\right.\nonumber
\\
&&\left.-2C_{g}\right)v_{\Gamma}. \label{Eq:VRGVGammaRM}
\end{eqnarray}
Here, we introduce a new parameter $C_{g}$ to characterize the
effective strength of disorder. For RSP and RM, it is
\begin{eqnarray}
C_g =\frac{v_{\Gamma}^{2}\Delta}{2\pi v_{1}v_{2}}.
\end{eqnarray}
For RVP, we have
\begin{eqnarray}
C_g = \frac{\left(v_{\Gamma1}^{2}+v_{\Gamma2}^{2}\right)
\Delta}{2\pi v_{1}v_{2}}.
\end{eqnarray}
The three coefficients $C_{0}^{A}$, $C_{1}^{A}$, and $C_{2}^{A}$
appearing in the coupled RG equations are
\begin{eqnarray}
C_{0}^{A}&=&\frac{1}{8\pi^3}\int_{-\infty}^{+\infty}dx
\int_{0}^{2\pi}d\theta\nonumber \\
&&\times\frac{x^2-\cos^2\theta-(v_{2}/v_{1})^{2}
\sin^2\theta}{\left(x^2+\cos^2\theta+(v_{2}/v_{1})^{2}
\sin^2\theta\right)^{2}} \mathcal{G}^{A}(x,\theta), \label{Eq:C0A}
\\
C_{1}^{A}&=&\frac{1}{8\pi^3}\int_{-\infty}^{+\infty} dx
\int_{0}^{2\pi}d\theta \nonumber
\\
&&\times\frac{-x^2+\cos^2\theta-(v_{2}/v_{1})^{2}\sin^2\theta}
{\left(x^2+\cos^2\theta+(v_{2}/v_{1})^{2}\sin^2\theta\right)^{2}}
\mathcal{G}^{A}(x,\theta), \label{Eq:C1A}
\\
C_{2}^{A}&=&\frac{1}{8\pi^3}\int_{-\infty}^{+\infty} dx
\int_{0}^{2\pi}d\theta\nonumber
\\
&&\times\frac{-x^2-\cos^2\theta+(v_{2}/v_{1})^{2}\sin^2\theta}
{\left(x^2+\cos^2\theta+(v_{2}/v_{1})^{2}\sin^2\theta\right)^{2}}
\mathcal{G}^{A}(x,\theta),\label{Eq:C2A}
\end{eqnarray}
where
\begin{eqnarray}
\mathcal{G}^{A}(x,\theta) = \frac{1}{\frac{N}{4v_2/v_{1}}
\sqrt{x^{2}+\cos^{2}\theta + \left(v_{2}/v_{1}\right)^{2}
\sin^{2}\theta}}. \label{Eq:GExpressionA}
\end{eqnarray}
The coefficients $C_{0}^{B}$, $C_{1}^{B}$, and $C_{2}^{B}$ are
\begin{eqnarray}
C_{0}^{B} &=& \frac{1}{8\pi^3}\int_{-\infty}^{+\infty}dx
\int_{0}^{2\pi}d\theta \nonumber \\
&&\times\frac{-x^2+\cos^2\theta + (v_{2}/v_{1})^{2}
\sin^2\theta}{\left(x^2+\cos^2\theta+(v_{2}/v_{1})^{2}
\sin^2\theta\right)^{2}}\mathcal{G}^{B}(x,\theta),
\label{Eq:C0B} \\
C_{1}^{B}&=&\frac{1}{8\pi^3}\int_{-\infty}^{+\infty}
dx\int_{0}^{2\pi}d\theta \nonumber \\
&&\times\frac{-x^2+\cos^2\theta-(v_{2}/v_{1})^{2}
\sin^2\theta}{\left(x^2 + \cos^2\theta + (v_{2}/v_{1})^{2}
\sin^2\theta\right)^{2}}\mathcal{G}^{B}(x,\theta), \label{Eq:C1B}
\\
C_{2}^{B} &=& \frac{1}{8\pi^3}\int_{-\infty}^{+\infty}dx
\int_{0}^{2\pi}d\theta \nonumber \\
&& \times \frac{-x^2-\cos^2\theta+(v_{2}/v_{1})^{2}
\sin^2\theta}{\left(x^2 + \cos^2\theta+(v_{2}/v_{1})^{2}
\sin^2\theta\right)^{2}} \mathcal{G}^{B}(x,\theta), \label{Eq:C2B}
\end{eqnarray}
with
\begin{eqnarray}
\mathcal{G}^{B}(x,\theta) = \frac{1}{\frac{1}{2\pi\alpha_{1}} +
\frac{N}{8v_2/v_{1}}\frac{\cos^2\theta+(v_{2}/v_{1})^{2}
\sin^2\theta}{\sqrt{x^2+\cos^2\theta+(v_{2}/v_{1})^{2}
\sin^2\theta}}}. \label{Eq:GExpressionB}
\end{eqnarray}
An effective parameter
\begin{eqnarray}
\alpha_{1} = \frac{e^2}{\epsilon v_{1}}
\end{eqnarray}
is defined to represent the Coulomb interaction strength. The
electric charge $e$ is not renormalized due to the absence of
logarithmic term in the polarization $\Pi^{B}$ \cite{Kotov12}, and
$\epsilon$ takes a constant value in any given sample. The value of
$\alpha_1$ is determined by the renormalization of velocity $v_1$.

The coupled flow equations can be simplified. According to
Eq.~(\ref{Eq:VRGVGammaRCP}), we know that
\begin{eqnarray}
v_{\Gamma} = v_{\Gamma 0}
\end{eqnarray}
is independent of $\ell$ for RSP. Thus we re-write $C_{g}$ as
\begin{eqnarray}
C_g = \frac{v_{\Gamma 0}^{2}\Delta}{2\pi v_{1}v_{2}}.
\end{eqnarray}
The flow equation for $C_{g}$ is given by
\begin{eqnarray}
\frac{dC_{g}}{d\ell} &=& \left(-2C_{0}^{A}-2C_{0}^{B} +
C_{1}^{A}+C_{1}^{B}+C_{2}^{A}+C_{2}^{B}\right.\nonumber \\
&&\left.+2C_{g}\right)C_{g}.
\end{eqnarray}
For RVP, from Eqs.~(\ref{Eq:VRGV1}), (\ref{Eq:VRGV2}),
(\ref{Eq:VRGVGammaRVP1}), and (\ref{Eq:VRGVGammaRVP2}), one gets
\begin{eqnarray}
\frac{d(v_{\Gamma1}/v_{1})}{d\ell}=0,\qquad
\frac{d(v_{\Gamma2}/v_{2})}{d\ell}=0,
\end{eqnarray}
which indicate that
\begin{eqnarray}
\frac{v_{\Gamma1}}{v_{1}}=\frac{v_{\Gamma10}}{v_{10}},\qquad
\frac{v_{\Gamma2}}{v_{2}}=\frac{v_{\Gamma20}}{v_{20}}.
\end{eqnarray}
Accordingly, $C_g$ now can be written as
\begin{eqnarray}
C_g = \frac{\Delta}{2\pi} \left(\frac{v_{\Gamma10}^{2}}{v_{10}^{2}}
\frac{v_{1}}{v_{2}}+\frac{v_{\Gamma20}^{2}}{v_{20}^{2}}
\frac{v_{2}}{v_{1}}\right).
\end{eqnarray}
The corresponding RG equation is
\begin{eqnarray}
\frac{dC_g}{d\ell} = \frac{\left(v_{\Gamma1}^{2} -
v_{\Gamma2}^{2}\right)\Delta}{2\pi v_{1}v_{2}}\left(-C_{1}^{A} -
C_{1}^{B}+C_{2}^{A} +C_{2}^{B}\right).
\end{eqnarray}

For RM, through Eqs.~(\ref{Eq:VRGV1}), (\ref{Eq:VRGV2}), and
(\ref{Eq:VRGVGammaRM}), we obtain the following flow equation
\begin{eqnarray}
\frac{d C_{g}}{d\ell} &=& \left(2C_{0}^{A} + 3C_{1}^{A} + 3C_{2}^{A}
+ 2C_{0}^{B} - C_{1}^{B} - C_{2}^{B}\right.\nonumber
\\
&&\left.-2C_{g}\right)C_{g}.
\end{eqnarray}

\section{Quantum critical phenomena \label{Sec:NumResults}}

In this section, we will solve the RG equations and then apply the
solutions to analyze the quantum critical phenomena. We adopt the
following steps: first, examine the low-energy behaviors induced
solely by the quantum critical fluctuation of excitonic order
parameter; second, introduce quenched disorder into the system and
study its interplay with the Yukawa coupling; finally, investigate
the impact of Coulomb interaction on the results.

Although the RG calculations are carried out at $T=0$, it is
possible to extract the $T$-dependence of observable quantities from
RG results. We can regard $k_B T$, where $k_B$ is Boltzmann
constant, as a free parameter that tunes the energy scale:
increasing (decreasing) $T$ amounts to increasing (decreasing) the
energy $\omega$. The dependence of observable quantities on $\omega$
and/or $T$ can be computed from the solutions of RG equations as
follows. One solves the flow equations at $T=0$ and gets the
$\ell$-dependence of model parameters, such as fermion velocities,
which leads to the $\ell$-dependence of various observable
quantities. On the basis of these results, one converts the
$\ell$-dependence of an observable quantity into the
$\omega$-dependence of the same quantity at $T=0$ by using the
transformation $\omega = \omega_0 e^{-\ell}$, where $\omega_0$ is
some high energy, or into the $T$-dependence of the same quantity by
using the transformation $T = T_0 e^{-\ell}$, where $T_0$ takes a
large value. For examples, the low-energy DOS $\rho(\omega)$ can be
directly obtained from $\rho(\ell)$, and the $T$-dependent specific
heat $C_{v}(T)$ can be obtained from $C_{v}(\ell)$. This approach
has been extensively employed to calculate the $\omega$- and/or
$T$-dependence of many observable quantities of Dirac/Weyl fermions
subject to the Coulomb interaction \cite{Goswami11, Hosur12, Moon13,
WangLiu14, Lai15, Jian15, Isobe16B, Cho16, WangLiuZhang17MWSM,
WangLiuZhangSemiDSM, Zhang17} and gapless nodal fermions coupled to
the nematic quantum fluctuation \cite{Huh08, Xu08, Fritz09, Wang11,
Liu12, She15, WangLiuZhang16NJP}.

\subsection{Non-Fermi liquid behavior induced by excitonic fluctuation
\label{SubSec:NumResOnlyEIFL}}

If 2D DSM is far from SM-EI transition, the ground state is a robust
SM. While the Coulomb interaction is long-ranged, it can only
produce normal FL behavior \cite{Kotov12, Gonzalez99, Hofmann14,
Hofmann15}. As the system approaches to the SM-EI QCP, the excitonic
fluctuation becomes stronger and eventually invalidates the FL
description at $T=0$. Now we illustrate how FL theory breaks down at
the QCP by analyzing the solutions of RG equations.

In the clean limit, the excitonic fluctuation leads to the following
RG equations
\begin{eqnarray}
\frac{dZ_{f}}{d\ell}&=&C_{0}^{A}Z_{f},
\\
\frac{dv_{1}}{d\ell}&=&\left(C_{0}^{A}-C_{1}^{A}\right)v_{1},
\\
\frac{dv_{2}}{d\ell}&=&\left(C_{0}^{A}-C_{2}^{A}\right)v_{2},
\\
\frac{d\left(v_{2}/v_{1}\right)}{d\ell}
&=&\left(C_{1}^{A}-C_{2}^{A}\right)\frac{v_{2}}{v_{1}}.
\end{eqnarray}
These equations will be solved in the isotropic and anisotropic
cases respectively.

\subsubsection{Isotropic limit}

We first consider the isotropic limit, i.e., $v_{1}=v_{2}=v$. In
this case, we have
\begin{eqnarray}
C_{0}^{A}=C_{1}^{A}=C_{2}^{A} = -\frac{2}{3\pi^{2}N}=-\eta^{A}.
\label{Eq:C1AC2AC3AIsotropic}
\end{eqnarray}
Accordingly, the RG equations can be simplified to
\begin{eqnarray}
\frac{dZ_{f}}{d\ell} &=& -\eta^{A}Z_{f}, \\
\frac{dv}{d\ell} &=& 0.
\end{eqnarray}
The velocity is a constant, i.e., $v = v_{0}$. Thus, the fermion
dispersion is unrenormalized, and the dynamical exponents is $z=1$
\cite{Herbut09}. The specific heat behaves as \cite{Herbut09}
\begin{eqnarray}
C_{v}(T)\sim T^{d/z}\sim T^{2}.
\end{eqnarray}
The residue is given by \cite{Herbut09}
\begin{eqnarray}
Z_{f} = Z_{f0}e^{-\eta^{A}\ell}=e^{-\eta^{A}\ell},
\end{eqnarray}
which flows to zero quickly in the limit $\ell\rightarrow \infty$.
$Z_{f}$ is connected to the real part of retarded self-energy
$\mathrm{Re}\Sigma^{R}(\omega)$ via the definition
\begin{eqnarray}
Z_{f} = \frac{1}{\left|1-\frac{\partial}{\partial\omega}
\mathrm{Re}\Sigma^{R}(\omega)\right|}.
\end{eqnarray}
Employing the transformation $\omega=\omega_{0}e^{-\ell}$, we get
the following expression
\begin{eqnarray}
\mathrm{Re}\Sigma^{R}(\omega)\sim\omega^{1-\eta^{A}}.
\end{eqnarray}
Using the Kramers-Kronig relation, we can easily obtain the
imaginary part
\begin{eqnarray}
\mathrm{Im}\Sigma^{R}(\omega)\sim\omega^{1-\eta^{A}},
\end{eqnarray}
which exhibits typical NFL behavior. The renormalized DOS depends on
$\omega$ as follows
\begin{eqnarray}
\rho(\omega)\sim\omega^{1+\eta^{A}}.
\end{eqnarray}

\subsubsection{Anisotropic case}

In the generic anisotropic case, namely $v_{1}\neq v_{2}$, we
integrate over variable $x$ in Eqs.~(\ref{Eq:C0A})-(\ref{Eq:C2A})
and find
\begin{eqnarray}
C_{0}^{A} &=&-\frac{v_{2}/v_{1}}{3\pi^3N}\int_{0}^{2\pi}d\theta
\frac{1}{\left(\cos^2\theta +
(v_{2}/v_{1})^{2}\sin^2\theta\right)}\nonumber \\
&=&-\frac{v_{2}/v_{1}}{3\pi^3N}\frac{2\pi}{v_{2}/v_{1}} = -\eta^{A},
\\
C_{1}^{A} &=& \frac{v_{2}/v_{1}}{3\pi^3 N}\int_{0}^{2\pi}d\theta
\frac{\cos^2\theta-3(v_{2}/v_{1})^{2}\sin^2\theta}{\left(\cos^2\theta
+ (v_{2}/v_{1})^{2}\sin^2\theta\right)^{2}}\nonumber
\\
&=&\frac{v_{2}/v_{1}}{3\pi^3 N}\left(-\frac{2 \pi}{v_{2}/v_{1}}
\right) = -\eta^{A}, \\
C_{2}^{A} &=&\frac{v_{2}/v_{1}}{3\pi^3 N}\int_{0}^{2\pi}d\theta
\frac{-3\cos^2\theta+(v_{2}/v_{1})^{2}\sin^2\theta}{\left(\cos^2
\theta + (v_{2}/v_{1})^{2}\sin^2\theta\right)^{2}} \nonumber \\
&=&\frac{v_{2}/v_{1}}{3\pi^3 N}\left(-\frac{2 \pi}{v_{2}/v_{1}}
\right)=-\eta^{A},
\end{eqnarray}
which are exactly the same as the isotropic case. Accordingly, the
RG equations for $v_{1}$ and $v_{2}$ are
\begin{eqnarray}
\frac{dv_{1}}{d\ell} = \frac{dv_{2}}{d\ell} = 0,
\end{eqnarray}
which implies that
\begin{eqnarray}
v_{1}=v_{10},\quad v_{2}=v_{20}.
\end{eqnarray}
Thus, the fermion velocities are not renormalized, and the
anisotropy is not changed by the Yukawa coupling. The low-energy
properties of specific heat $C_{v}(T)$, residue $Z_{f}$, fermion
damping rate $|\mathrm{Im}\Sigma^{R}(\omega)|$, and DOS
$\rho(\omega)$ are the same as those obtained in the isotropic case.

\subsection{Excitonic fluctuation and disorder}

We then include disorder and examine how it affects the above
results. Now the coupled RG equations of $Z_{f}$, $v_{1}$, $v_{2}$,
and $v_{2}/v_{1}$ are
\begin{eqnarray}
\frac{dZ_{f}}{d\ell} &=& \left(C_{0}^{A}-C_{g}\right)Z_{f} =
-\left(\eta^{A}+C_{g}\right)Z_{f}, \label{Eq:VRGZfQFDis}
\\
\frac{dv_{1}}{d\ell} &=& \left(C_{0}^{A}-C_{1}^{A}-C_{g}\right)v_{1}
= -C_{g}v_{1}, \label{Eq:VRGV1QFDis}
\\
\frac{dv_{2}}{d\ell} &=& \left(C_{0}^{A}-C_{2}^{A} -
C_{g}\right)v_{2}=-C_{g}v_{2}, \label{Eq:VRGV2QFDis}
\\
\frac{d\left(v_{2}/v_{1}\right)}{d\ell} &=& \left(C_{1}^{A} -
C_{2}^{A}\right)\frac{v_{2}}{v_{1}}=0. \label{Eq:VRGVRatioQFDis}
\end{eqnarray}

For RSP, $C_{g}$ satisfies
\begin{eqnarray}
\frac{dC_{g}}{d\ell} = 2C_{g}^{2}, \label{Eq:VRGCgRCPQFDis}
\end{eqnarray}
whose solution is
\begin{eqnarray}
C_{g} = \frac{C_{g0}}{1-2C_{g0}\ell}. \label{Eq:SolutionCgRCPQFDis}
\end{eqnarray}
It is clear that this $C_{g}$ diverges as $\ell \rightarrow
\ell_{c}$, where $\ell_{c} = 1/2C_{g0}$. Substituting
Eq.~(\ref{Eq:SolutionCgRCPQFDis}) into
Eqs.~(\ref{Eq:VRGZfQFDis})-(\ref{Eq:VRGV2QFDis}), we obtain
\begin{eqnarray}
Z_{f}&=&e^{-\eta^{A}\ell}\sqrt{1-2C_{g0}\ell},
\\
v_{1}&=&v_{10}\sqrt{1-2C_{g0}\ell},
\\
v_{2}&=&v_{20}\sqrt{1-2C_{g0}\ell}.
\end{eqnarray}
We can see that, $Z_{f}$, $v_1$, and $v_2$ all flow to zero as $\ell
\rightarrow \ell_{c}$. Such singular behaviors are generally
believed to indicate the instability of the system: RSP drives the
system into a disorder-dominated CDM. The characteristic feature of
CDM is that, the fermions acquire a finite disorder scattering rate
\begin{eqnarray}
\gamma_{\mathrm{imp}} = \left|\mathrm{Im}\Sigma^{R}(0)\right|.
\end{eqnarray}
In the meantime, the zero-energy DOS $\rho(0)$ also becomes finite,
being a function of $\gamma_{\mathrm{imp}}$. According to the
calculations given in Refs.\cite{WangLiu14, WangLiuZhang16NJP}, the
specific heat displays a linear-in-$T$ behavior, namely
\begin{eqnarray}
C_{v}(T)\sim T.
\end{eqnarray}
The NFL quantum critical state realized in the clean limit is turned
into a CDM once RSP is added to the system, even when RSP is very
weak. The fermion damping effect, the low-energy DOS, and the
specific heat of CDM phase are all distinct from those of the NFL
phase.

For RVP, the RG equation for $C_{g}$ is
\begin{eqnarray}
\frac{dC_g}{d\ell}=\frac{\left(v_{\Gamma1}^{2}-v_{\Gamma2}^{2}
\right)\Delta}{2\pi v_{1}v_{2}} \left(-C_{1}^{A}+C_{2}^{A}\right) =
0, \label{Eq:VRGCgRGPQFDis}
\end{eqnarray}
implying that
\begin{eqnarray}
C_{g} = C_{g0}. \label{Eq:SolutionCgRGPQFDis}
\end{eqnarray}
Substituting Eq.~(\ref{Eq:SolutionCgRGPQFDis}) into
Eqs.~(\ref{Eq:VRGZfQFDis})-(\ref{Eq:VRGV2QFDis}) yields
\begin{eqnarray}
Z_{f}&=&e^{-\left(\eta^{A}+C_{g0}\right)\ell}, \\
v_{1}&=&v_{10}e^{-C_{g0}\ell}, \\
v_{2}&=&v_{20}e^{-C_{g0}\ell}.
\end{eqnarray}
The real and imaginary parts of retarded fermion self-energy are
\begin{eqnarray}
\mathrm{Re}\Sigma^{R}(\omega)\sim \omega^{1 -
\left(\eta^{A}+C_{g0}\right)}, \\
\mathrm{Im}\Sigma^{R}(\omega)\sim \omega^{1 -
\left(\eta^{A}+C_{g0}\right)},
\end{eqnarray}
which are still NFL-like behaviors. Comparing to the clean limit,
$Z_{f}$ approaches to zero more quickly and the fermion damping
becomes stronger. The velocity $v$ goes to zero rapidly with growing
$\ell$, thus the fermion dispersion is substantially altered. In
addition, the dynamical exponent $z$ becomes $z = 1+C_{g0}$. It is
easy to find that, the specific heat is
\begin{eqnarray}
C_{v}(T)\sim T^{d/z}\sim T^{2/(1+C_{g0})},
\end{eqnarray}
and the low-energy DOS is
\begin{eqnarray}
\rho(\omega)\sim \omega^{(1-C_{g0})/(1+C_{g0})+\eta^{A}}.
\end{eqnarray}
An apparent conclusion is that both DOS and specific heat are
enhanced by RVP at low energies.

For RM, the RG equation for $C_{g}$ becomes
\begin{eqnarray}
\frac{dC_{g}}{d\ell} = -8\eta^{A}C_{g}-2C_{g}^{2}.
\label{Eq:VRGCgRMQFDis}
\end{eqnarray}
Its solution is
\begin{eqnarray}
C_{g}(\ell) = \frac{4\eta^{A}C_{g0}}{\left(C_{g0} + 4\eta^{A}\right)
e^{8\eta^{A}\ell}-C_{g0}}, \label{Eq:SolutionCgRMQFDis}
\end{eqnarray}
which vanishes in the limit $\ell \rightarrow \infty$. Substituting
Eq.~(\ref{Eq:SolutionCgRMQFDis}) into
Eqs.~(\ref{Eq:VRGZfQFDis})-(\ref{Eq:VRGV2QFDis}), we get
\begin{eqnarray}
Z_{f} &=& e^{-\eta^{A}\ell}\sqrt{\frac{4\eta^{A}}{C_{g0} + 4\eta^{A}
- C_{g0}e^{-8\eta^{A}\ell}}}, \label{Eq:VRGZfCgRMQFDis}
\\
v_{1} &=& v_{10}\sqrt{\frac{4\eta^{A}}{C_{g0} + 4\eta^{A} -
C_{g0}e^{-8\eta^{A}\ell}}}, \label{Eq:VRGV1CgRMQFDis}
\\
v_{2} &=& v_{20}\sqrt{\frac{4\eta^{A}}{C_{g0} + 4\eta^{A} -
C_{g0}e^{-8\eta^{A}\ell}}}. \label{Eq:VRGV2CgRMQFDis}
\end{eqnarray}
In the low-energy regime, the residue still behaves as $Z_{f} \sim
e^{-\eta^{A}\ell}$. From the $\ell$-dependence of $Z_{f}$, we obtain
\begin{eqnarray}
\mathrm{Re}\Sigma^{R}(\omega) &\sim& \omega^{1-\eta^{A}},
\\
\mathrm{Im}\Sigma^{R}(\omega)&\sim&\omega^{1-\eta^{A}},
\end{eqnarray}
which are the same as the clean case. As shown by
Eqs.~(\ref{Eq:VRGV1CgRMQFDis}) and (\ref{Eq:VRGV2CgRMQFDis}),
$v_{1}$ and $v_{2}$ approach to finite values in the lowest energy
limit. Accordingly, the fermion DOS still exhibits the behavior
$\rho(\omega)\sim\omega^{1+\eta^{A}}$, and the specific heat is
still of the form $C_{v}(T)\sim T^{2}$. We thus see that RM does not
qualitatively change the low-energy properties of observable
quantities.

The above RG results indicate that, the low-energy properties of the
SM-EI QCP depend heavily on the disorder type. Such properties can
be experimentally probed by measuring observable quantities, such as
DOS and specific heat. However, we should remember that the
long-range Coulomb interaction is entirely ignored in the above RG
analysis. This might miss important quantum many-body effects. In
the next subsection, we will study whether or not the above results
are substantially altered when the Coulomb interaction is
incorporated.

\subsection{Interplay of three kinds of interaction}

We now analyze the physical consequence of the interplay of all the
three kinds of interaction, first in the isotropic limit and then in
the more generic anisotropic case. We will see that the Coulomb
interaction tends to suppress the fermion velocity anisotropy.

\begin{figure}[htbp]
\center
\includegraphics[width=3.38in]{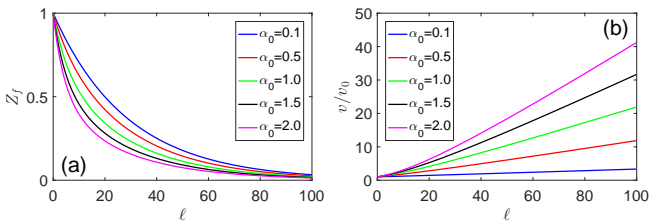}
\caption{Flowing behavior of $Z_{f}$ and $v$ caused by excitonic
fluctuation and Coulomb interaction. In this and all the subsequent
figures, we assume $N=2$ in numerical calculations.}
\label{Fig:VRGIsoClean}
\end{figure}

\begin{figure}[htbp]
\center
\includegraphics[width=3.38in]{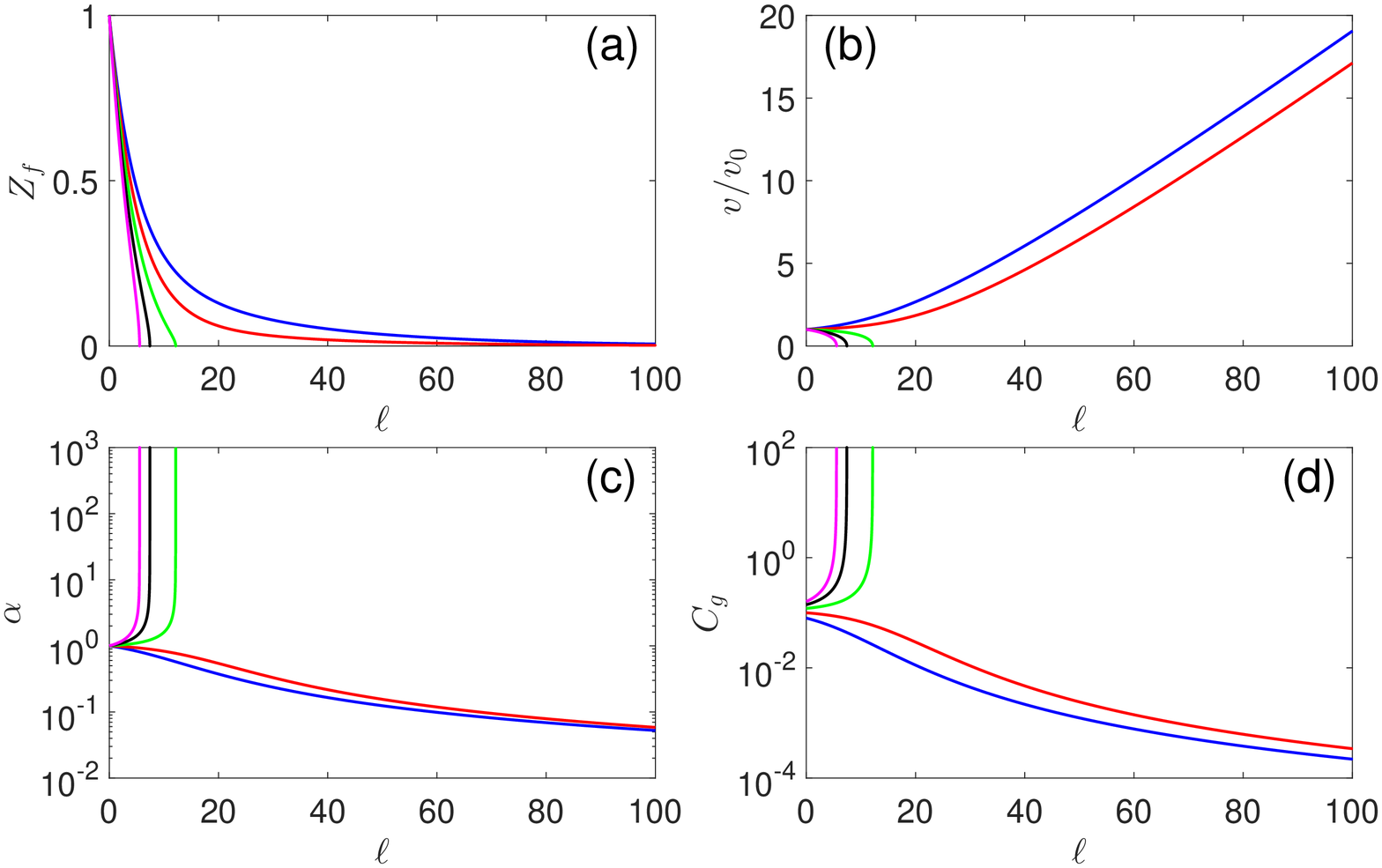}
\caption{Flowing behavior of $Z_{f}$, $v$, $\alpha$, and $C_{g}$
caused by excitonic fluctuation, Coulomb interaction, and RSP. Blue,
red, green, black, and magenta lines correspond to $C_{g0} = 0.08,
0.1, 0.12, 0.14, 0.16$. Here, $\alpha_{10}=1.0$.
\label{Fig:VRGIsoRSP}}
\end{figure}

\subsubsection{Isotropic limit}

In the isotropic limit with $v_{1} = v_{2} = v$, the RG equations
for $Z_{f}$ and $v$ are
\begin{eqnarray}
\frac{dZ_{f}}{d\ell} &=& \left(-\eta^{A}+C_{0}^{B}-C_{g}\right)
Z_{f}, \label{Eq:VRGZfQFDisCoulombIso}
\\
\frac{dv}{d\ell}&=&\left(C^{B}-C_{g}\right)v.
\label{Eq:VRGVQFDisCoulombIso}
\end{eqnarray}
Here, $C^{B} = C_{0}^{B}-C_{1}^{B} = C_{0}-C_{2}^{B}$, in which
\begin{eqnarray}
C_{0}^{B} &=& \frac{4}{N\pi^2}\left[2-\frac{1}{\lambda}\pi +
\frac{2-\lambda^{2}}{\lambda}f(\lambda)\right], \\
C_{1,2}^{B} &=& \frac{4}{N\pi^2}\left[1-\frac{1}{\lambda}
\frac{\pi}{2}+\frac{1-\lambda^{2}}{\lambda}f(\lambda)\right].
\end{eqnarray}
The variable $\lambda$ is $\lambda = N\pi\alpha/4$, and the function
$f(\lambda)$ is
\begin{eqnarray}
f(\lambda)=\left\{
\begin{array}{ll}
\frac{1}{\sqrt{1-\lambda^2}}\arccos\left(\lambda\right) & \lambda<1
\\
\\
\frac{1}{\sqrt{\lambda^2-1}}\mathrm{arccosh}\left(\lambda\right) &
\lambda>1
\\
\\
1 & \lambda=1.
\end{array}
\right.
\end{eqnarray}

In the clean limit, $Z_{f}$ and $v$ flow as follows
\begin{eqnarray}
\frac{dZ_{f}}{d\ell} &=& \left(-\eta^{A}+C_{0}^{B}\right)Z_{f},
\label{Eq:VRGZfQFCoulombIso} \\
\frac{dv}{d\ell} &=& C^{B}v. \label{Eq:VRGVQFCoulombIso}
\end{eqnarray}
The numerical solutions are shown in Fig.~\ref{Fig:VRGIsoClean}. The
velocity $v$ increases as the energy is lowered. The Coulomb
interaction is marginally irrelevant since its strength parameter
$\alpha = e^{2}/v\epsilon$ flows to zero slowly in the lowest energy
limit. Both $C_{0}^{B}$ and $C^{B}$ vanish as $\alpha \rightarrow
0$. The velocity renormalization produces logarithmic-like
correction to the temperature or energy dependence of some
observable quantities, including specific heat and compressibility
\cite{Kotov12}. The singular renormalization of fermion velocities
has been observed by various experimental tools \cite{Elias11,
Siegel11, Chae12, Yu13}. At low energies, $C_{0}^{B}$ is much
smaller than $\eta^{A}$. Thus, the Coulomb interaction only slightly
alters the low-energy behavior of $Z_{f}$ induced by the excitonic
fluctuation.

\begin{figure}[htbp]
\center
\includegraphics[width=2.38in]{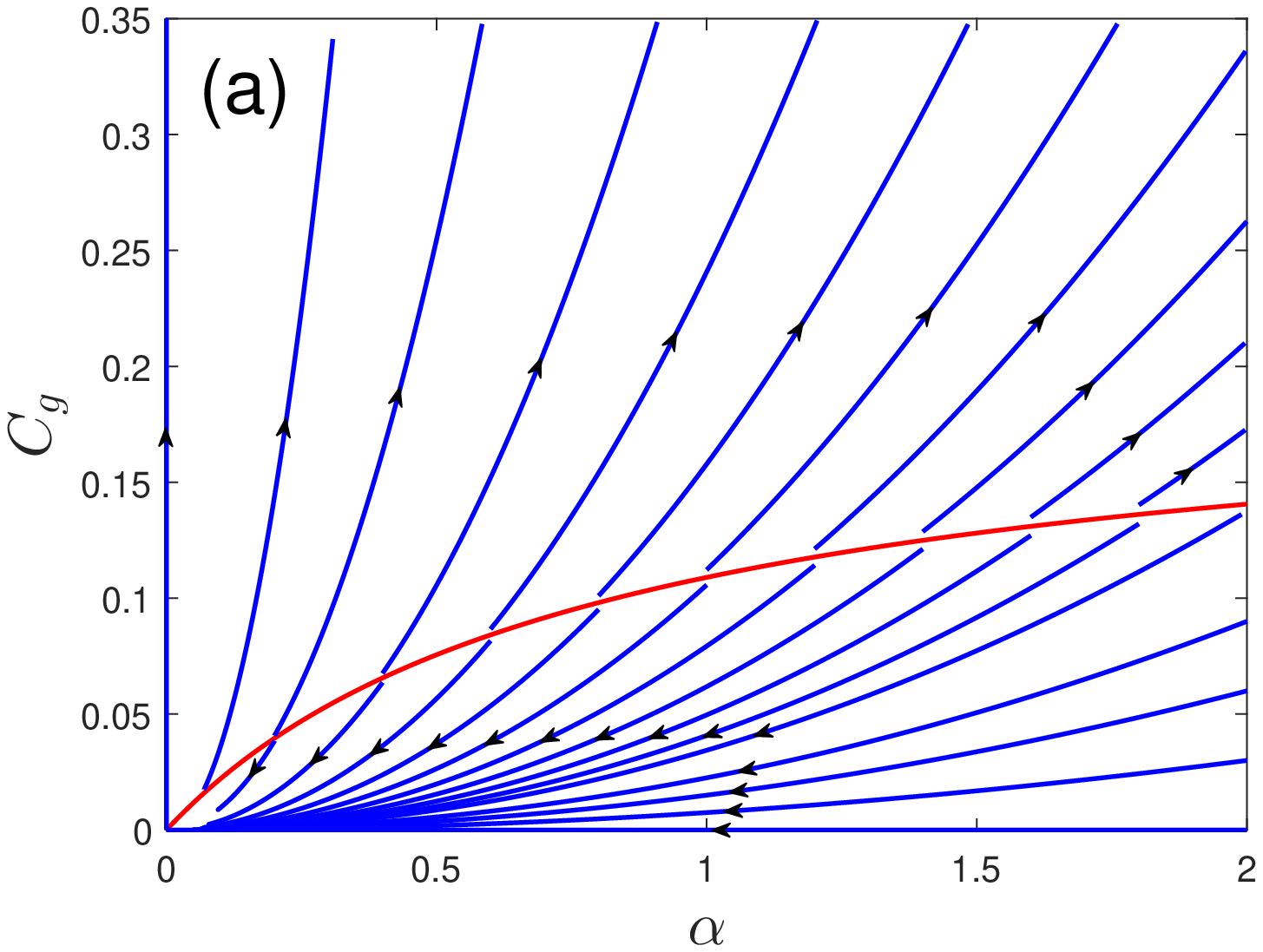}
\includegraphics[width=2.38in]{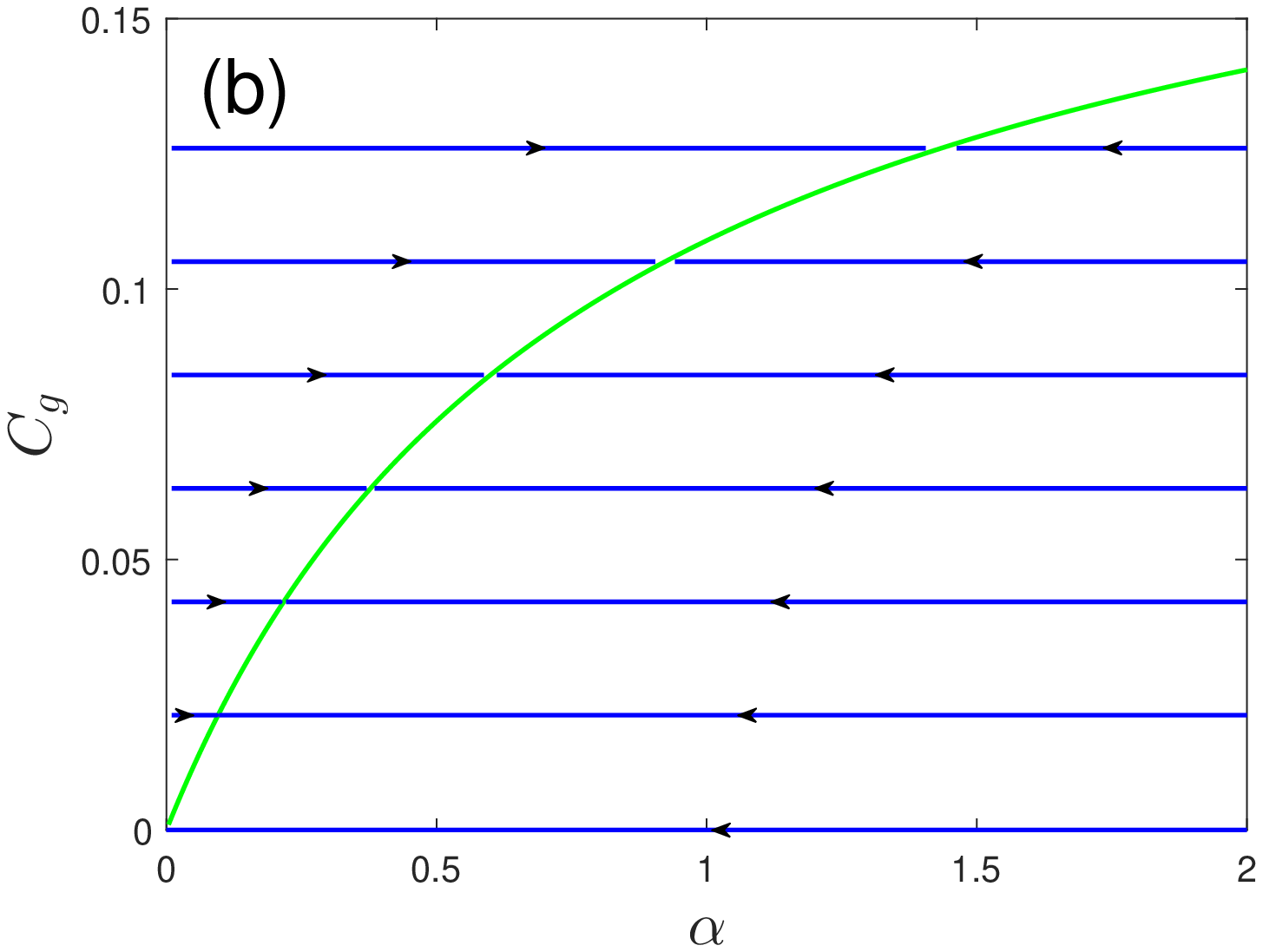}
\includegraphics[width=2.38in]{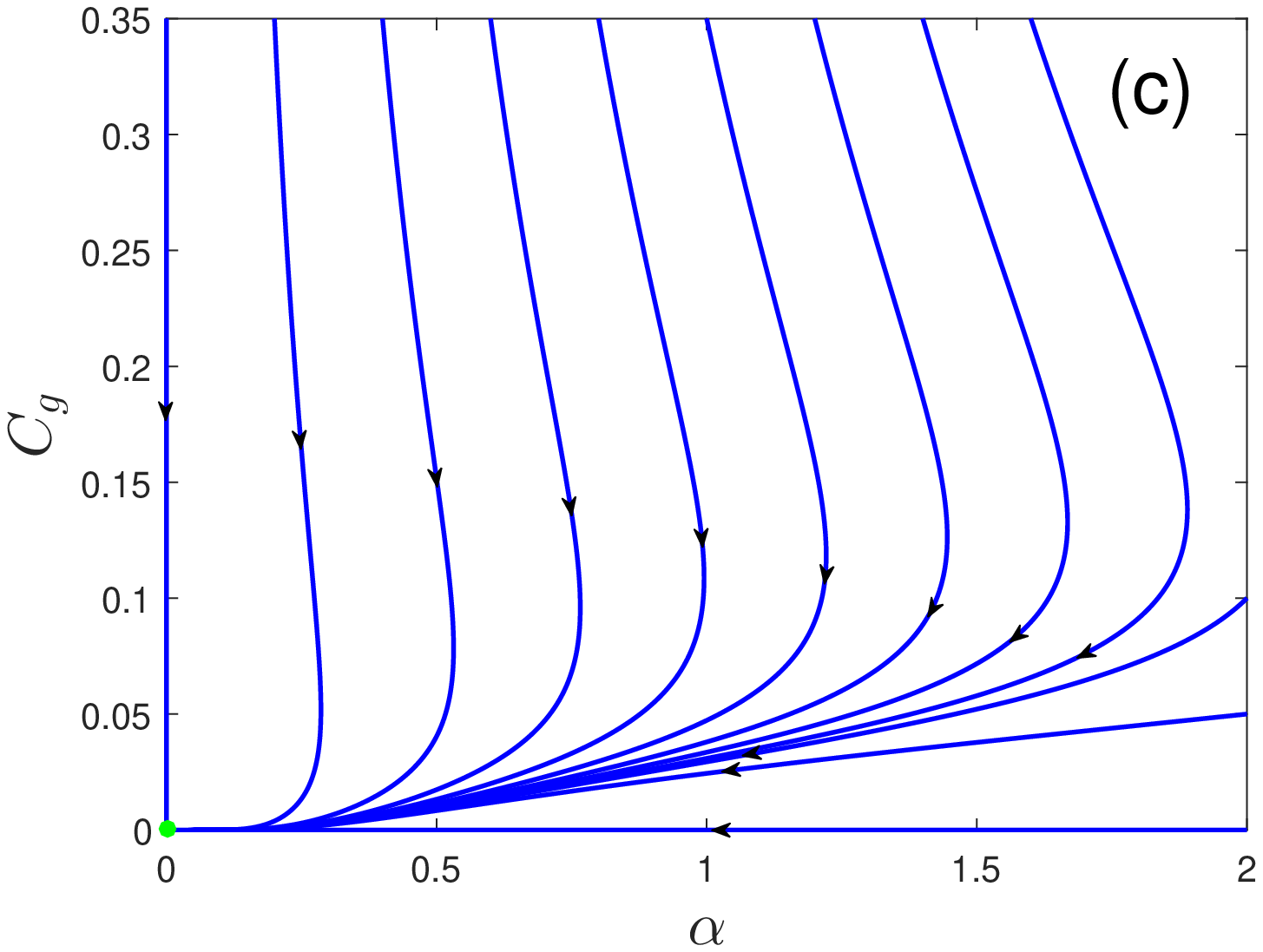}
\caption{Flowing diagrams on the $\alpha$-$C_{g}$ plane. Result for
RSP is in (a), RVP in (b), and RM in (c). \label{Fig:FlowDiagrams}}
\end{figure}

For RSP, the RG equation of $C_{g}$ is
\begin{eqnarray}
\frac{dC_{g}}{d\ell} = \left(-2C^{B}+2C_{g}\right)C_{g}.
\end{eqnarray}
For a given $\alpha_0$, there exists a critical value
$C^B(\alpha_0)$. The system exhibits entirely different low-energy
properties when $C_{g0}$ is greater and smaller than
$C^{B}(\alpha_{0})$. To illustrate this, we show the
$\ell$-dependence of $Z_f$, $v$, $\alpha$, and $C_g$ in
Fig.~\ref{Fig:VRGIsoRSP}. If $C_{g0} < C^{B}(\alpha_{0})$, $Z_f$,
$\alpha$, and $C_{g}$ all flow to zero as $\ell \rightarrow \infty$,
but $v$ increases with growing $\ell$. These results indicate that
weak RSP is suppressed by the Coulomb interaction. If
$C_{g0}>C^{B}(\alpha_{0})$, both $C_{g}$ and $\alpha$ formally
diverge at some finite energy scale, whereas both $Z_f$ and $v$
decrease rapidly down to zero at the same energy scale. Thus, strong
RSP still drives a NFL-to-CDM transition. As can be seen from the
flow diagram presented in Fig.~\ref{Fig:FlowDiagrams}(a), the
$(\alpha, C_{g})$ plane is divided by the critical line $C_{g0} =
C^{B}(\alpha_{0})$ into two distinct phases: the NFL phase and the
CDM phase.

\begin{figure}[htbp]
\center
\includegraphics[width=3.3in]{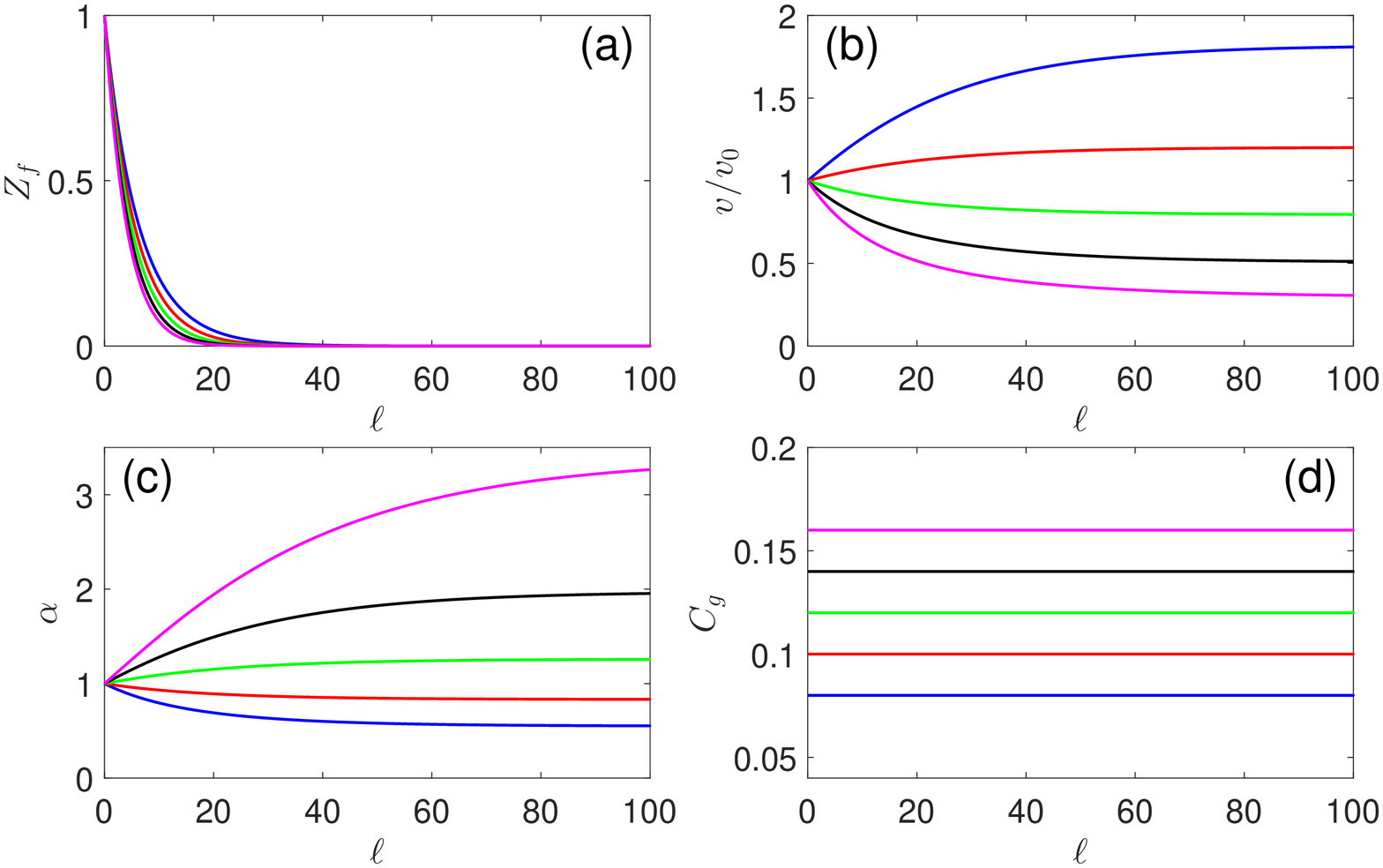}
\caption{Flowing behavior of $Z_{f}$, $v$, $\alpha$, and $C_{g}$
caused by excitonic fluctuation, Coulomb interaction, and RVP. Blue,
red, green, black, and magenta lines correspond to $C_{g0} = 0.08,
0.1, 0.12, 0.14, 0.16$. Here, $\alpha_{10}=1.0$.
\label{Fig:VRGIsoRVP}}
\end{figure}

For RVP, the $\ell$-dependence of $Z_f$, $v$, $\alpha$, and $C_g$
are shown in Fig.~\ref{Fig:VRGIsoRVP}. The parameter $C_{g}$ does
not flow at all, namely
\begin{eqnarray}
\frac{dC_{g}}{d\ell} = 0.
\end{eqnarray}
We fix $C_g$ at a constant: $C_{g} = C_{g0}$. For a given $C_{g0}$,
$v$ approaches to a constant value $v^{*}$ in the zero energy limit.
The value of $v^{*}$ is obtained from
\begin{eqnarray}
C^{B}(\alpha^{*}) = C_{g0},
\end{eqnarray}
where $\alpha^{*} = e^{2}/v^{*}\epsilon$. RG analysis indicates that
the system always flows to a stable infrared fixed point for any two
given initial values of $\alpha$ and $C_g$. Connecting all of these
fixed points forms a critical line on the $\alpha$-$C_g$ plane, as
shown in Fig.~\ref{Fig:FlowDiagrams}(b). Near the critical line, the
specific heat behaves as
\begin{eqnarray}
C_{v}(T)\sim \frac{1}{{v^{*}}^{2}}T^{2} \sim T^{2}.
\end{eqnarray}
The residue is
\begin{eqnarray}
Z_{f} &\sim& e^{\left(-\eta^{A} + C_{0}^{B}(\alpha^{*}) -
C_{g0}\right)\ell}\nonumber \\
&\sim& e^{\left(-\eta^{A}+C_{1}^{B}(\alpha^{*})\right)\ell},
\end{eqnarray}
where $C_{1}^{B}(\alpha^{*})$ is negative. This $Z_{f}$ flows to
zero more quickly than that induced purely by excitonic fluctuation.
The retarded fermion self-energy is
\begin{eqnarray}
\mathrm{Re}\Sigma^{R}(\omega)&\sim& \omega^{1-\left(\eta^{A}
- C_{1}^{B}(\alpha^{*})\right)}, \\
\mathrm{Im}\Sigma^{R}(\omega)&\sim& \omega^{1 -
\left(\eta^{A}-C_{1}^{B}(\alpha^{*})\right)}.
\end{eqnarray}
The DOS takes the form
\begin{eqnarray}
\rho(\omega) \sim \omega^{1 + \eta^{A} - C_{1}^{B}(\alpha^{*})}.
\end{eqnarray}

\begin{figure}[htbp]
\center
\includegraphics[width=3.3in]{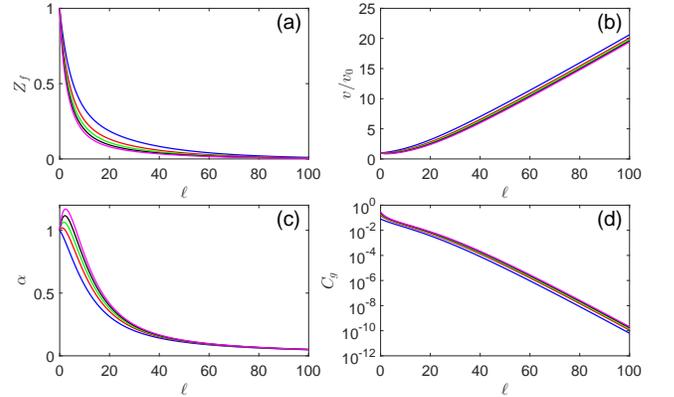}
\caption{Flowing behavior of $Z_{f}$, $v$, $\alpha$, and $C_{g}$
caused by excitonic fluctuation, Coulomb interaction, and RM. Blue,
red, green, black, and magenta lines correspond to $C_{g0}=0.08,
0.15, 0.2, 0.25, 0.3$. Here, $\alpha_{10}=1.0$.
\label{Fig:VRGIsoRM}}
\end{figure}

For RM, the RG equation for $C_{g}$ is given by
\begin{eqnarray}
\frac{d C_{g}}{d\ell} = \left(-8\eta^{A}+2C^{B}-2C_{g}\right)C_{g}.
\end{eqnarray}
The numerical results are plotted in Fig.~\ref{Fig:VRGIsoRM}. We
observe that $C_{g}$ always approaches to zero quickly, which
indicates that RM is irrelevant in the low-energy regime. The
Coulomb interaction is marginally irrelevant and leads to singular
renormalization of fermion velocity. Accordingly, the DOS and
specific heat are
\begin{eqnarray}
\rho(\omega) &\sim& \frac{\omega^{1 + \eta^{A}
}}{\ln^{2}(\omega_{0}/\omega)}, \\
C_{v}(T) &\sim& \frac{T^{2}}{\ln^{2}(T_{0}/T)}.
\end{eqnarray}
In the presence of RM, the two parameters $(\alpha, C_{g})$ always
flow to the stable infrared fixed point $(0,0)$.

\begin{table*}[htbp]
\caption{A summary of low-energy or low-temperature behaviors of
some characteristic quantities caused by all the possible
combination of the three types of interaction. QFEO stands for the
quantum fluctuation of excitonic order parameter. CI represents the
Coulomb interaction. We choose to display the $\ell$-dependent
quasiparticle residue $Z_{f}(\ell)$, the fermion damping rate
$\mathrm{Im}\Sigma^{R}(\omega)$, the DOS $\rho(\omega)$, and the
specific heat $C_{v}(T)$. The definitions of all the notations are
given in the main text. \label{Table:SummaryObQuant}}
\begin{center}
\begin{tabular}{|c|c|c|c|c|c|}
\hline\hline   \multicolumn{2}{|c|}{Interaction}  &  $Z_{f}(\ell)$ &
$\mathrm{Im}\Sigma^{R}(\omega)$ & $\rho(\omega)$ &  $C_{v}(T)$
\\
\hline      \multicolumn{2}{|c|}{QFEO} &
\tabincell{c}{$e^{-\eta^{A}\ell}$ \\ \cite{Herbut09}} &
\tabincell{c}{$\omega^{1-\eta^{A}}$ \\ \cite{Herbut09} } &
$\omega^{1+\eta^{A}}$ & \tabincell{c}{ $T^{2}$ \\ \cite{Herbut09}}
\\
\hline   \multicolumn{2}{|c|}{QFEO+RSP}   &
$e^{-\eta^{A}\ell}\sqrt{1-2C_{g0}\ell}$  & $\gamma_{\mathrm{imp}}$ &
$ \gamma_{\mathrm{imp}}\ln(v\Lambda/\gamma_{\mathrm{imp}})$ &
$\rho(0)T$
\\
\hline   \multicolumn{2}{|c|}{QFEO+RVP}   &
$e^{-\left(\eta^{A}+C_{g0}\right)\ell}$ &
$\omega^{1-\left(\eta^{A}+C_{g0}\right)}$ & $
\;\omega^{(1-C_{g0})/(1+C_{g0})+\eta^{A}}\;$ & $T^{2/(1+C_{g0})}$
\\
\hline   \multicolumn{2}{|c|}{QFEO+RM}   & $e^{-\eta^{A}\ell}$  &
$\omega^{1-\eta^{A}}$ & $\omega^{1+\eta^{A}}$ &  $T^{2}$
\\
\hline   \multicolumn{2}{|c|}{QFEO+CI}   & $e^{-\eta^{A}\ell} $  &
$\omega^{1-\eta^{A}}$ &
$\omega^{1+\eta^{A}}/\ln^2(\omega_{0}/\omega)$ & $\;
T^{2}/\ln^{2}\left(T_{0}/T\right)\; $
\\
\hline   \multirow{2}{*}{QFEO+CI+RSP}   & $C_{g0}<
C_{B}(\alpha_{0})$   & $e^{-\eta^{A}\ell}$ & $\omega^{1-\eta^{A}}$ &
$\omega^{1+\eta^{A}}/\ln^{2}(\omega_{0}/\omega)$ &
$T^{2}/\ln^{2}(T_{0}/T)$
\\
\cline{2-6} & $C_{g0}> C_{B}(\alpha_{0})$  & $\lim_{\ell\rightarrow
l_{c}}Z_{f}(\ell)\rightarrow0$ & $\gamma_{\mathrm{imp}}$ & $
\gamma_{\mathrm{imp}}\ln(v\Lambda/\gamma_{\mathrm{imp}})$ &
$\rho(0)T$
\\
\hline   \multicolumn{2}{|c|}{QFEO+CI+RVP}   &
$e^{\left(-\eta^{A}+C_{1}^{B}(\alpha^{*})\right)\ell}$ &
$\omega^{1-\left(\eta^{A}-C_{1}^{B}(\alpha^{*})\right)}$ &
$\omega^{1+\eta^{A}-C_{1}^{B}(\alpha^{*})}$ & $T^{2}$
\\
\hline   \multicolumn{2}{|c|}{QFEO+CI+RM}   & $e^{-\eta^{A}\ell}$ &
$\omega^{1-\eta^{A}}$ &
$\omega^{1+\eta^{A}}/\ln^{2}(\omega_{0}/\omega)$ &
$T^{2}/\ln^{2}\left(T_{0}/T\right)$
\\
\hline  \multicolumn{2}{|c|}{CI} &
\tabincell{c}{$\;\lim_{\ell\rightarrow\infty}Z_{f}(\ell)\rightarrow
\mathrm{Const.}\;$ \\ \cite{Kotov12, Gonzalez99, Hofmann14, WangLiu14}} &
\tabincell{c}{$\omega/\ln^{2}(\omega_{0}/\omega)$ \\ \cite{Kotov12, WangLiu14}}
& \tabincell{c}{$\omega/\ln^{2}(\omega_{0}/\omega)$ \\
\cite{Kotov12, WangLiu14} } & \tabincell{c}{$T^{2}/\ln^{2}(T_{0}/T)$ \\
\cite{Kotov12, WangLiu14}}
\\
\hline   \multirow{2}{*}{CI+RSP}   & $C_{g0}< C_{B}(\alpha_{0})$ &
\tabincell{c}{$\lim_{\ell\rightarrow\infty}Z_{f}(\ell)\rightarrow \mathrm{Const.}$ \\
\cite{WangLiu14}} &
\tabincell{c}{$\omega/\ln^{2}(\omega_{0}/\omega)$ \\
\cite{WangLiu14}} &
\tabincell{c}{$\omega/\ln^{2}(\omega_{0}/\omega)$ \\
\cite{WangLiu14, Stauber05}}  & \tabincell{c}{$T^{2}/\ln^{2}(T_{0}/T)$\\
\cite{WangLiu14, Stauber05}}
\\
\cline{2-6} & $C_{g0} > C_{B}(\alpha_{0})$  &
\tabincell{c}{$\lim_{\ell\rightarrow l_{c}}Z_{f}(\ell)\rightarrow0$
\\ \cite{WangLiu14} } & \tabincell{c}{$\gamma_{\mathrm{imp}}$ \\
\cite{Sarma11}} &
\tabincell{c}{$\gamma_{\mathrm{imp}}\ln(v\Lambda/\gamma_{\mathrm{imp}})$
\\ \cite{Sarma11}}& \tabincell{c}{$\rho(0)T$  \\ \cite{Sarma11}}
\\
\hline   \multicolumn{2}{|c|}{CI+RVP}   &
\tabincell{c}{$e^{C_{1}^{B}(\alpha^{*})\ell}$ \\ \cite{WangLiu14}} &
\tabincell{c}{$\omega^{1+C_{1}^{B}(\alpha^{*})}$ \\
\cite{WangLiu14}} & \tabincell{c}{$\omega^{1-C_{1}^{B}(\alpha^{*})}$
\\ \cite{WangLiu14}} & \tabincell{c}{$T^{2}$ \\ \cite{WangLiu14,
Stauber05, Herbut08, Vafek08}}
\\
\hline   \multicolumn{2}{|c|}{CI+RM}   &
\tabincell{c}{$e^{C_{1}^{B}(\alpha^{*})\ell}$ \\ \cite{WangLiu14}} &
\tabincell{c}{$\omega^{1+C_{1}^{B}(\alpha^{*})}$ \\
\cite{WangLiu14}} & \tabincell{c}{$\omega^{1-C_{1}^{B}(\alpha^{*})}$
\\ \cite{WangLiu14}} & \tabincell{c}{$T^{2}$ \\ \cite{WangLiu14,
Stauber05, Herbut08, Vafek08}}
\\
\hline \hline
\end{tabular}
\end{center}
\end{table*}

We now compare the quantum critical phenomena to the physical
properties of the SM phase. Deep in the SM phase, the excitonic
fluctuation can be completely ignored. The low-energy behavior is
governed by the interplay of Coulomb interaction and disorder, which
has already been extensively investigated \cite{WangLiu14, Ye98,
Ye99, Stauber05, Herbut08, Vafek08, Foster08}. When the Coulomb
interaction and RSP are both present, the system is a normal FL if
RSP is weak, but is turned into a CDM phase by strong RSP. Thus,
increasing the effective strength of RSP drives a FL-CDM phase
transition. In the SM-EI quantum critical regime, increasing the
effective strength of RSP leads to a NFL-CDM transition. If RM is
added to the system, it is irrelevant around the SM-EI QCP, but is
marginal and results in a stable critical line on the $\alpha$-$C_g$
plane deep in the SM phase. In contrast, RVP produces the same
qualitative low-energy behaviors in the SM phase and around the
SM-EI QCP.

We learn from the above analysis that, even if 2D DSM has a gapless
SM ground state, the fluctuation of excitonic order parameter gives
rise to observable effects at finite $T$ and/or $\omega$. The
quantum critical regime can be distinguished from the pure SM phase
by measuring the $\omega$-dependence of fermion damping rate and/or
the $T$-dependence of specific heat.

To provide a complete analysis of the quantum critical phenomena, we
summarize in Table~\ref{Table:SummaryObQuant} the low-energy
properties induced by all the possible combinations of three types
of interaction. The quantities presented in
Table~\ref{Table:SummaryObQuant} include the residue $Z_f$, damping
rate $\mathrm{Im}\Sigma^{R}(\omega)$, fermion DOS $\rho(\omega)$,
and specific heat $C_{v}(T)$. We can see that distinct interactions
affect each other significantly. The critical phenomena cannot be
reliably determined if their mutual influence is not carefully
handled.

\subsubsection{Anisotropic case}

\begin{figure}[htbp]
\center
\includegraphics[width=3.3in]{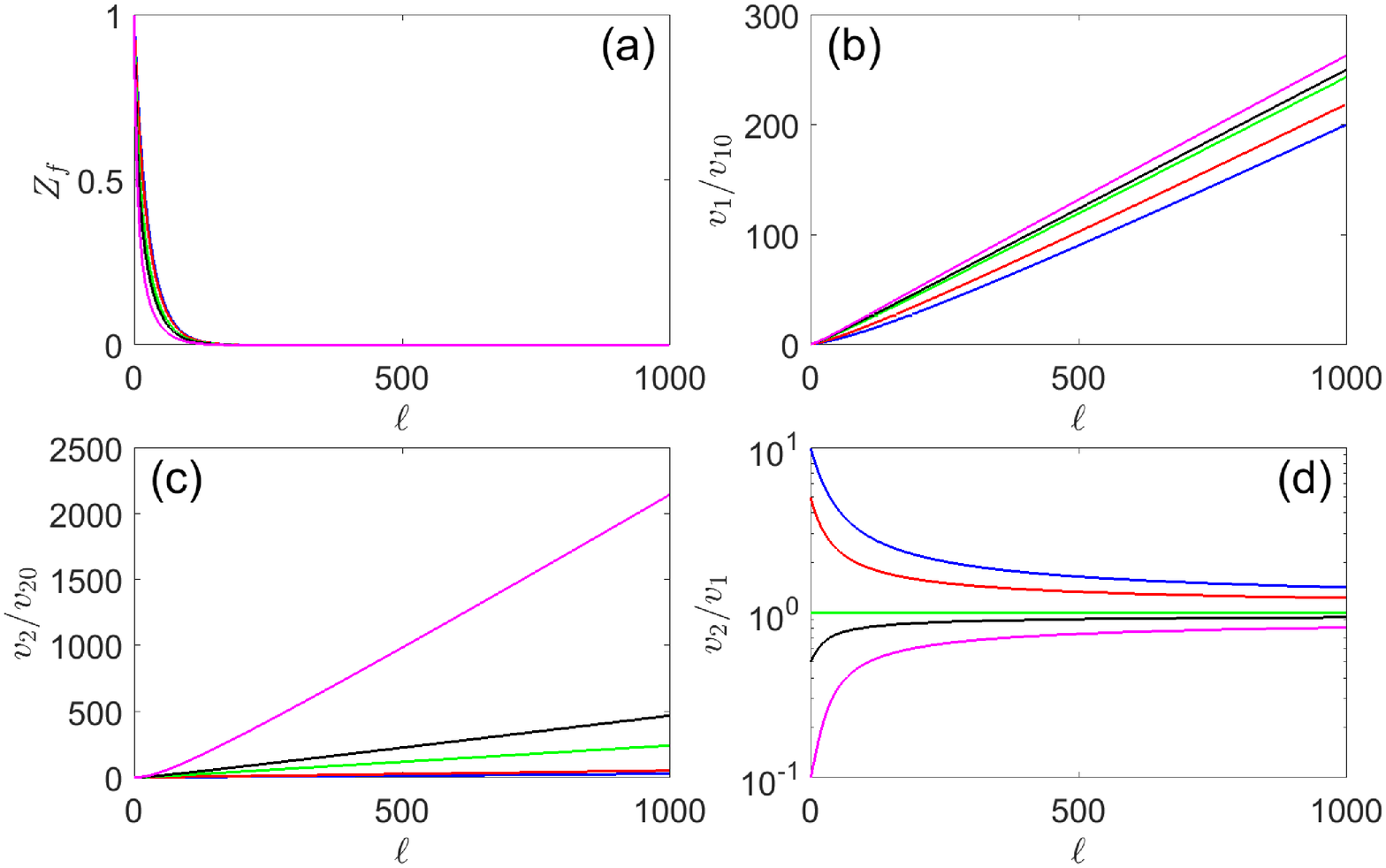}
\caption{Flowing behavior of $Z_{f}$, $v_{1}$, $v_{2}$, and
$v_{2}/v_{1}$ caused by excitonic fluctuation and Coulomb
interaction in the anisotropic case. Blue, red, green, black, and
magenta lines correspond to $v_{20}/v_{10}=10, 5, 1, 0.5, 0.1$. We
choose $\alpha_{10} = 1.0$. As $\ell \rightarrow \infty$, the system
flows to the isotropic limit. \label{Fig:VRGAniClean}}
\end{figure}

For different values of fermion velocity ratio, the running
behaviors of $Z_{f}$, $v_{1}$, $v_{2}$, and $v_{2}/v_{1}$ obtained
in the clean limit are plotted in
Figs.~\ref{Fig:VRGAniClean}(a)-(d), respectively. Firstly, $Z_{f}$
flows to zero very quickly, implying the violation of FL
description. This is essentially induced by the excitonic quantum
fluctuation, because the Coulomb interaction by itself would yield a
finite $Z_f$. Secondly, the two fermion velocities $v_1$ and $v_2$
both increase as the energy is lowered, whereas the velocity ratio
$v_2/v_1$ flows to unity in the lowest energy limit. Remember that
the excitonic quantum fluctuation does not renormalize fermion
velocities at all, as illustrated in
Sec.~\ref{SubSec:NumResOnlyEIFL}. It is clear that the
renormalization of $v_{1}$ and $v_{2}$ are mainly determined by the
Coulomb interaction. These results indicate that both excitonic
fluctuation and Coulomb interaction are important in the low-energy
region.

\begin{figure}[htbp]
\center
\includegraphics[width=3.3in]{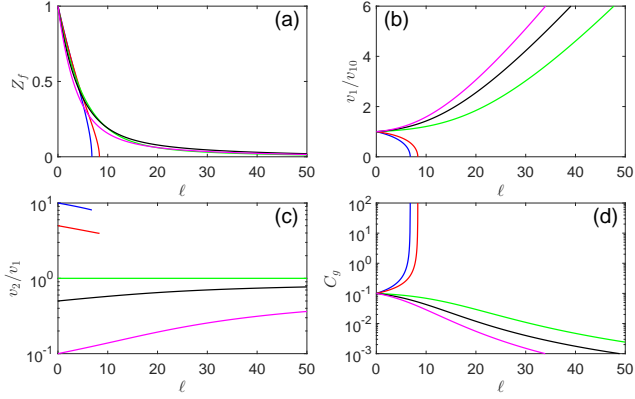}
\caption{Flowing behavior of $Z_{f}$, $v_{1}$, $v_{2}/v_{1}$, and
$C_{g}$ caused by excitonic fluctuation, Coulomb interaction, and
RSP. Blue, red, green, black, and magenta lines correspond to
$v_{20}/v_{10} = 10, 5, 1, 0.5, 0.1$. Here, $\alpha_{10}=1.0$ and
$C_{g0}=0.1$. \label{Fig:VRGAniRSP}}
\end{figure}

\begin{figure}[htbp]
\center
\includegraphics[width=3.3in]{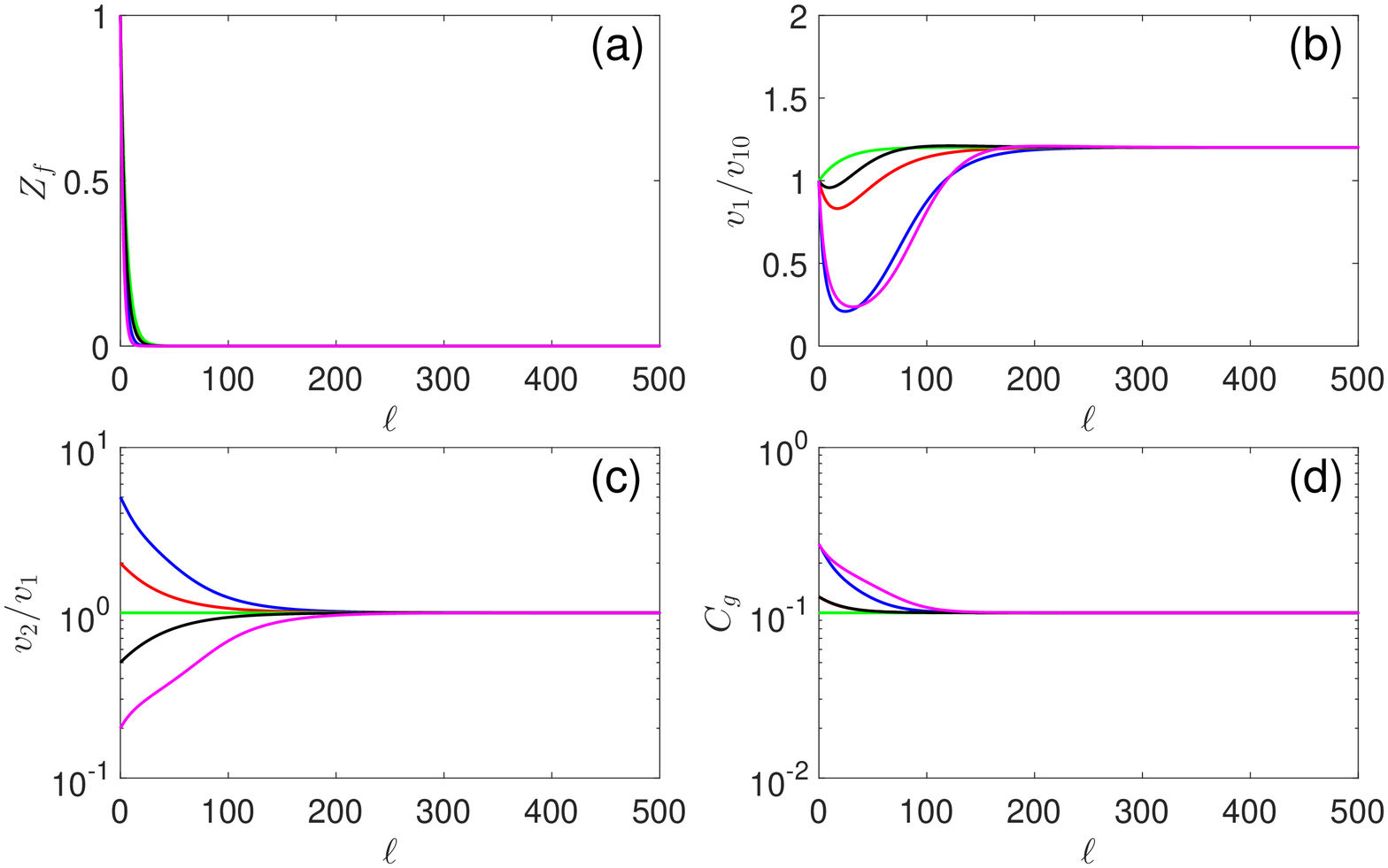}
\caption{Flowing behavior of $Z_{f}$, $v_{1}$, $v_{2}/v_{1}$, and
$C_{g}$ caused by excitonic fluctuation, Coulomb interaction, and
RVP. Blue, red, green, black, and magenta lines correspond to
$v_{20}/v_{10}=5, 2, 1, 0.5, 0.2$. Here, $\alpha_{10}=1.0$,
$\Delta/2\pi = 0.05$, $v_{\Gamma10}/v_{10} = 1$, and
$v_{\Gamma20}/v_{20} = 1$. \label{Fig:VRGAniRVP}}
\end{figure}

\begin{figure}[htbp]
\center
\includegraphics[width=3.3in]{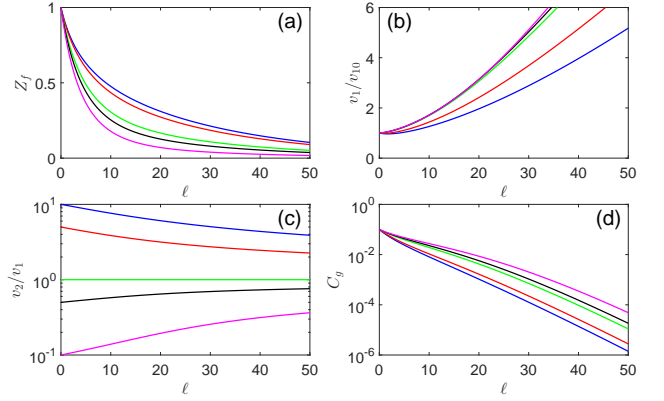}
\caption{Flowing behavior of $Z_{f}$, $v_{1}$, $v_{2}/v_{1}$, and
$C_{g}$ caused by excitonic fluctuation, Coulomb interaction, and
RM. Blue, red, green, black, and magenta lines correspond to
$v_{20}/v_{10}=10, 5, 1, 0.5, 0.1$. Here, $\alpha_{10}=1.0$ and
$C_{g0}=0.1$. \label{Fig:VRGAniRM}}
\end{figure}

After including three types of disorder, we find that the system
still flows to the isotropic limit in the zero energy limit. The
numerical results obtained in the cases of RSP, RVP, and RM are
presented in Fig.~\ref{Fig:VRGAniRSP}, Fig.~\ref{Fig:VRGAniRVP}, and
Fig.~\ref{Fig:VRGAniRM}, respectively.

First, we consider the case of RSP. As shown in
Fig.~\ref{Fig:VRGAniRSP}, for given values of $\alpha_{10}$ and
$C_{g0}$, $C_{g}$ becomes divergent at some finite energy scale if
the bare velocity ratio $v_{20}/v_{10}$ exceeds a critical value.
Both $Z_{f}$ and fermion velocities flow to zero at the same energy
scale. The anisotropy is suppressed, but the ratio does not flow to
the isotropic limit. If the bare value $v_{20}/v_{10}$ is small,
$C_{g0}$ flows to zero quickly as the energy is lowered. Meanwhile,
the fermion velocities increase, and the ratio
$v_{2}/v_{1}\rightarrow 1$. Apparently, the isotropic limit is
mainly driven by the Coulomb interaction. The residue $Z_{f}$ still
vanishes, owing to the excitonic fluctuation. For given values of
$\alpha_{10}$ and $C_{g0}$, varying the velocity ratio
$v_{20}/v_{10}$ leads to QPT between CDM phase and NFL phase.

In the case of RVP, we show the evolution of $Z_{f}$, $v_{1}$,
$v_{2}/v_{1}$, and $C_{g}$ in Fig.~\ref{Fig:VRGAniRVP}. Comparing to
the clean limit, the ratio $v_{2}/v_{1}$ approaches to unity more
quickly. This should be attributed to the fact that the Coulomb
interaction strength $\alpha$ flows to certain finite value in the
presence of RVP but vanishes in the clean limit. Therefore, the
suppression of velocity anisotropy is more significant once RVP is
introduced.

We finally turn to the impact of RM. According to
Fig.~\ref{Fig:VRGAniRM}, the disorder parameter $C_{g}$ of RM always
flows to zero quickly with decreasing energy. The low-energy
behaviors of $Z_{f}$ and $v_{1}$ are nearly the same as those
obtained in the clean limit, and the velocity ratio $v_{2}/v_{1}
\rightarrow 1$ as the energy is lowered down to zero.

\section{Summary and Discussion \label{Sec:SummaryDiscuss}}

In summary, we have presented a systematic study of the quantum
critical phenomena around the SM-EI QCP in 2D DSM. The Yukawa
coupling between Dirac fermions and excitonic quantum fluctuation,
the long-range Coulomb interaction, and the disorder scattering are
treated on equal footing, focusing on their mutual influence and the
consequent low-energy properties of the quantum critical regime. We
first studied the influence of quantum critical fluctuation of
excitonic order parameter, and showed that it invalidates the FL
description. We further demonstrated that, adding RSP always drives
a NFL-to-CDM transition, and adding RVP further reinforces the NFL
behaviors. Nevertheless, adding RM does not change the qualitative
results obtained in the clean limit. Once Coulomb interaction is
also incorporated, the above results are altered. In particular, the
NFL state is protected by the Coulomb interaction for weak RSP, but
is eventually replaced by CDM state if RSP is strong enough. When
RVP or RM coexist with excitonic fluctuation and Coulomb
interaction, the system is in a NFL state. To characterize the NFL
and CDM phases, we have calculated several quantities, including the
residue, damping rate, fermion DOS, and specific heat. The predicted
quantum critical phenomena can be directly probed by experiments.

The results obtained in this paper might be applied to judge whether
or not a 2D DSM is close to the SM-EI QCP. Deep in the gapless SM
phase, the properties of the system are determined by the
combination of Coulomb interaction and disorder. As the system
approaches the SM-EI QCP, i.e., $\alpha \rightarrow \alpha_c$, the
excitonic quantum fluctuation becomes progressively more important,
driving the system to enter into the quantum critical regime. Even
when the zero-$T$ ground state is gapless, the system could exhibit
nontrivial quantum critical behaviors in the $\omega$- and/or
$T$-dependence of observable quantities, as illustrated in
Fig.~\ref{Fig:PhaseEIQCP} and Table~\ref{Table:SummaryObQuant}.

We finally give a brief remark on the existence of the excitonic QCP
in realistic graphene. For a 2D DSM, all the previous analytical and
numerical calculations \cite{CastroNetoPhysics09, Khveshchenko01,
Gorbar02, Khveshchenko04, Liu09, Khveshchenko09, Gamayun10, Sabio10,
Zhang11, Liu11, WangLiu11A, WangLiu11B, WangLiu12, Popovici13,
WangLiu14, Gonzalez15, Carrington16, Sharma17, Xiao17, Carrington18,
Gamayun09, WangJianhui11, Katanin16, Gonzalez10, Gonzalez12,
Drut09A, Drut09B, Drut09C, Armour10, Armour11, Buividovich12,
Ulybyshev13, Smith14, Juan12, Kotikov16, Gonzalez14, Braguta16,
Xiao18} have confirmed that an excitonic gap is generated only when
$\alpha > \alpha_c$, where $\alpha_{c}$ is a nonzero critical value.
Recent theoretical studies revealed that the physical value of
$\alpha$ in suspendend graphene is not far from the critical value
$\alpha_{c}$ \cite{Ulybyshev13, Carrington18}. The system would
become even closer to the excitonic QCP when strain is applied
\cite{Tang15, Sharma17, Xiao17}. The organic conductor
$\alpha$-(BEDT-TTF)$_{2}$I$_{3}$, an anisotropic 2D DSM, may also be
close to the excitonic QCP \cite{Hirata17}. The theoretical results
obtained in this work could be utilized to explore the quantum
critical phenomena around the putative excitonic QCP in 2D DSM
materials.

\section*{ACKNOWLEDGEMENTS}

We thank Jing Wang and Peng-Lu Zhao for helpful discussions. We
acknowledge the financial support by the National Natural Science
Foundation of China under Grants 11275100, 11504379, and 11574285.
X.Y.P. also acknowledges the support by the K. C. Wong Magna
Foundation in Ningbo University. J.R.W. is partly supported by the
Natural Science Foundation of Anhui Province under Grant
1608085MA19.

\appendix

\section{Polarization functions}

We now calculate the polarization functions caused by the
particle-hole collective excitations. There are two polarization
functions, corresponding to the dynamical screening effects of the
quantum critical fluctuation of excitonic order parameter and the
long-range Coulomb interaction, respectively.

\subsection{Polarization function for excitonic fluctuation}

For the quantum excitonic fluctuation, the polarization function is
defined as
\begin{eqnarray}
\Pi^{A}(\Omega,\mathbf{q}) &=& N\int \frac{d\omega}{2\pi} \frac{d^2
\mathbf{k}}{(2\pi)^2}\mathrm{Tr}
\left[G_{0}(\omega,\mathbf{k})\right.\nonumber
\\
&&\left.\times G_{0}(\omega+\Omega,\mathbf{k}+\mathbf{q})\right].
\label{Eq:PolarizationADef}
\end{eqnarray}
Substituting the free fermion propagator into
Eq.~(\ref{Eq:PolarizationADef}), we obtain
\begin{eqnarray}
\Pi^{A}(\Omega,\mathbf{q})=-\frac{4N}{v_{1}v_{2}}\int \frac{d^3
k}{(2\pi)^3}\frac{k\cdot(k+q)}{k^2(k+q)^2},
\end{eqnarray}
where $k=(\omega,\mathbf{k})$. Here, we have employed the following
transformations
\begin{eqnarray}
v_{1}k_{1}\rightarrow k_{1}, \quad v_{2}k_{2}\rightarrow k_{2},
\quad v_{1}q_{1}\rightarrow q_{1}, \quad v_{2}q_{2}\rightarrow
q_{2}. \nonumber \\
\end{eqnarray}
Using the Feynman parametrization formula
\begin{eqnarray}
\frac{1}{AB}=\int_{0}^{1}dx \frac{1}{[Ax+(1-xB)]^2},
\label{Eq:FeynmanParaFormula}
\end{eqnarray}
one gets
\begin{eqnarray}
\Pi^{A}(\Omega,\mathbf{q}) &=& -\frac{4N}{v_{1}v_{2}}
\int_{0}^{1}dx\int\frac{d^3 k}{(2\pi)^3}\nonumber
\\
&&\times\frac{k\cdot(k+q)}{\left[(k+xq)^{2}+x(1-x)q^{2}\right]}.
\end{eqnarray}
Let $k+xq\rightarrow k$, $\Pi_{A}$ can be further written as
\begin{eqnarray}
\Pi^{A}(\Omega,\mathbf{q}) &=& -\frac{4N}{v_{1}v_{2}} \int^1_0
dx\left\{\int \frac{d^3 k}{(2\pi)^3} \frac{k^{2}}{[k^2 +
x(1-x)q^2]^2}\right.\nonumber \\
&& \left. -\int \frac{d^3 k}{(2\pi)^3}
\frac{x(1-x)q^2}{[k^2+x(1-x)q^2]^2}\right\}.
\end{eqnarray}
Performing integration over $k$ by using the standard formula of
dimensional regularization
\begin{eqnarray}
\int \frac{d^dk}{(2\pi)^d}\frac{1}{(k^2+\Delta)^n} &=&
\frac{1}{(4\pi)^{d/2}}\frac{\Gamma(n-\frac{d}{2})}{\Gamma(n)}
\frac{1}{\Delta^{n-\frac{d}{2}}}, \label{Eq:FormulaDimReGA}
\\
\int \frac{d^dk}{(2\pi)^d}\frac{k^2}{(k^2+\Delta)^n} &=&
\frac{1}{(4\pi)^{d/2}}\frac{d}{2}\frac{\Gamma(n -
\frac{d}{2}-1)}{\Gamma(n)}\frac{1}{\Delta^{n-\frac{d}{2}-1}},
\label{Eq:FormulaDimReGB} \nonumber \\
\end{eqnarray}
we find that
\begin{eqnarray}
\Pi^{A}(\Omega,\mathbf{q}) &=& \frac{2N}{v_{1}v_{2}\pi}
\sqrt{q^{2}}\int^1_0 dx \sqrt{x(1-x)}\nonumber \\
&=&\frac{N}{4v_{1}v_{2}}\sqrt{\Omega^2+q_{1}^{2}+q_{2}^{2}}.
\end{eqnarray}
By taking $q_{1}\rightarrow v_{1}q_{1}$ and $q_{2}\rightarrow
v_{2}q_{2}$, we get
\begin{eqnarray}
\Pi^{A}(\Omega,\mathbf{q}) = \frac{N}{4v_1 v_2} \sqrt{\Omega^2 +
v_{1}^{2}q_{1}^{2}+v_{2}^{2}q_{2}^{2}}. \label{Eq:PolarizationAResultApp}
\end{eqnarray}

\subsection{Polarization function for Coulomb interaction}

For the Coulomb interaction, the polarization function is given by
\begin{eqnarray}
\Pi^{B}(\Omega,\mathbf{q}) &=& -N\int\frac{d\omega}{2\pi}
\frac{d^2\mathbf{k}}{(2\pi)^2}\mathrm{Tr} \left[\gamma_{0}
G_{0}(\omega,\mathbf{k})\gamma_{0}\right.\nonumber
\\
&&\left.\times G_{0}(\omega+\Omega,\mathbf{k}+\mathbf{q})\right].
\label{Eq:PolarizationBDef}
\end{eqnarray}
Substituting Eq.~(\ref{Eq:FermionPropagator}) into
Eq.~(\ref{Eq:PolarizationBDef}) leads to
\begin{eqnarray}
\Pi^{B}(\Omega,\mathbf{q})=\frac{4N}{v_{1}v_{2}}\int \frac{d^3
k}{(2\pi)^3}\frac{2k_{0}(k_{0}+q_{0}) -
k\cdot(k+q)}{k^2(k+q)^2}.\nonumber \\
\end{eqnarray}
Making use of the Feynman parametrization formula
Eq.~(\ref{Eq:FeynmanParaFormula}), along with the transformation
$k+xq \rightarrow k$, we recast the above expression as
\begin{eqnarray}
\Pi^{B}(\Omega,\mathbf{q}) &=& \frac{4N}{v_{1}v_{2}} \int^1_0
dx\left\{\int \frac{d^3 k}{(2\pi)^3} \frac{-k^{2}/3}{[k^2 +
x(1-x)q^2]^2} \right.\nonumber \\
&&\left. +\int\frac{d^3 k}{(2\pi)^3}\frac{x(1-x) \left(q^2 -
2q_{0}^{2}\right)}{[k^2+x(1-x)q^2]^2}\right\}.
\end{eqnarray}
Repeating the calculational steps that lead to
Eq.~(\ref{Eq:PolarizationAResultApp}), we finally obtain
\begin{eqnarray}
\Pi^{B}(\Omega,\mathbf{q}) = \frac{N}{8v_1 v_2} \frac{v_{1}^2
q_{1}^{2} + v_{2}^2q_{2}^{2}}{\sqrt{\Omega^2 + v_{1}^{2}q_{1}^{2} +
v_{2}^{2}q_{2}^{2}}}.
\end{eqnarray}

\begin{widetext}
\section{Fermion self-energy}

The fermion self-energy corrections come from three sorts of
interaction, namely the Yukawa coupling, Coulomb interaction, and
disorder scattering. The former two interactions are inelastic, and
the third one is elastic. We now calculate them in order.

\subsection{Contribution from Yukawa coupling}

The fermion self-energy induced by the Yukawa coupling takes the
form
\begin{eqnarray}
\Sigma^{A}(\omega,\mathbf{k}) &=& \int'\frac{d\Omega}{2\pi}
\frac{d^2{\mathbf{q}}}{(2\pi)^{2}} G_{0}(\Omega+\omega,\mathbf{q} +
\mathbf{k})D^{A}(\Omega,\mathbf{q})\nonumber
\\
&=& -\int'\frac{d\Omega}{2\pi}\frac{d^2{\mathbf{q}}}{(2\pi)^{2}}
\frac{\left[-i(\Omega+\omega)\gamma_{0}+v_{1}(q_{1}+k_{1})
\gamma_{1}+v_{2}(q_{2}+k_{2})\gamma_{2}\right]} {\left[(\Omega +
\omega)^{2}+v_{1}^{2}(q_{1}+k_{1})^{2} + v_{2}^{2}(q_{2} +
k_{2})^{2}\right]} D^{A}(\Omega,\mathbf{q}).
\end{eqnarray}
This self-energy can be expanded in powers of $i\omega$,
$v_{1}k_{1}$, and $v_{2}k_{2}$. To the leading order, we get
\begin{eqnarray}
\Sigma^{A}(\omega,\mathbf{k}) &=& i\omega\gamma_{0}\int'
\frac{d\Omega}{2\pi}\frac{d^2{\mathbf{q}}}{(2\pi)^{2}}
\frac{-\Omega^{2}+v_{1}^{2}q_{1}^{2}+v_{2}^{2}
q_{2}^{2}}{\left(\Omega^{2}+v_{1}^{2}q_{1}^{2} +
v_{2}^{2}q_{2}^{2}\right)^{2}}\frac{1}{\frac{N}{4v_1v_2}
\sqrt{\Omega^2+v_{1}^{2}q_{1}^{2}+v_{2}^{2}q_{2}^{2}}}
\\
&&-v_{1}k_{1}\gamma_{1}\int'\frac{d\Omega}{2\pi}
\frac{d^2{\mathbf{q}}}{(2\pi)^{2}}\frac{\Omega^{2} -
v_{1}^{2}q_{1}^{2}+v_{2}^{2}q_{2}^{2}}{\left(\Omega^{2} +
v_{1}^{2}q_{1}^{2}+v_{2}^{2}q_{2}^{2}\right)^{2}}
\frac{1}{\frac{N}{4v_1v_2}
\sqrt{\Omega^2+v_{1}^{2}q_{1}^{2}+v_{2}^{2}q_{2}^{2}}}
\\
&&-v_{2}k_{2}\gamma_{2}\int'\frac{d\Omega}{2\pi}
\frac{d^2{\mathbf{q}}}{(2\pi)^{2}}\frac{\Omega^{2} +
v_{1}^{2}q_{1}^{2}-v_{2}^{2}q_{2}^{2}}{\left(\Omega^{2} +
v_{1}^{2}q_{1}^{2}+v_{2}^{2}q_{2}^{2}\right)^{2}}
\frac{1}{\frac{N}{4v_1v_2}\sqrt{\Omega^2 +
v_{1}^{2}q_{1}^{2}+v_{2}^{2}q_{2}^{2}}}.
\end{eqnarray}
To carry out RG calculation, we choose to integrate over the
integral variables within the range
\begin{eqnarray}
\int'\frac{d\Omega}{2\pi}\frac{d^2{\mathbf{q}}}{(2\pi)^{2}} =
\frac{1}{8\pi^{3}}\int_{-\infty}^{+\infty}d\Omega\int_{0}^{2\pi}
d\theta\int_{b\Lambda}^{\Lambda}d|\mathbf{q}|\left|\mathbf{q}\right|,
\label{Eq:IntegrationRange}
\end{eqnarray}
where $b=e^{-\ell}$. It is then easy to obtain
\begin{eqnarray}
\Sigma^{A}(\omega,\mathbf{k}) = \left(-i\omega\gamma_{0}C_{0}^{A} +
v_{1}k_{1}\gamma_{1}C_{1}^{A} + v_{2}k_{2}\gamma_{2}
C_{2}^{A}\right)\ell.
\end{eqnarray}
The expressions of $C_{i}^{A}$ are given by
Eqs.~(\ref{Eq:C0A})-(\ref{Eq:GExpressionA}).

\subsection{Contribution from Coulomb interaction}

The fermion self-energy induced by the Coulomb interaction is
\begin{eqnarray}
\Sigma^{B}(\omega,\mathbf{k}) &=& -\int'\frac{d\Omega}{2\pi}
\frac{d^2{\mathbf{q}}}{(2\pi)^{2}}\gamma_{0} G(\Omega +
\omega,\mathbf{q}+\mathbf{k})\gamma_{0}D^{B}(\Omega,\mathbf{q})
\nonumber \\
&=&\int'\frac{d\Omega}{2\pi}\frac{d^2{\mathbf{q}}}{(2\pi)^{2}}
\gamma_{0}\frac{\left[-i(\Omega+\omega)\gamma_{0}+v_{1}(q_{1} +
k_{1})\gamma_{1}+v_{2}(q_{2}+k_{2})\gamma_{2}\right]}{\left[(\Omega+\omega)^{2}
+ v_{1}^{2}(q_{1}+k_{1})^{2} +v_{2}^{2}(q_{2}+k_{2})^{2}\right]}
\gamma_{0}D^{B}(\Omega,\mathbf{q}).
\end{eqnarray}
To the leading order of small energy/momenta expansion, $\Sigma^{B}$
can be approximately written as
\begin{eqnarray}
\Sigma^{B}(\omega,\mathbf{k}) &=& -i\omega\gamma_{0}\int'
\frac{d\Omega}{2\pi}\frac{d^2{\mathbf{q}}}{(2\pi)^{2}}
\frac{-\Omega^{2}+v_{1}^{2}q_{1}^{2}+v_{2}^{2}
q_{2}^{2}}{\left(\Omega^{2}+v_{1}^{2}q_{1}^{2}+v_{2}^{2}q_{2}^{2}
\right)^{2}}\frac{1}{\frac{|\mathbf{q}|}{\frac{2\pi
e^{2}}{\epsilon}}+\frac{N}{8v_1 v_2} \frac{v_{1}^{2}q_{1}^{2} +
v_{2}^{2}q_{2}^{2}}{\sqrt{\Omega^2+v_{1}^{2}q_{1}^{2} +
v_{2}^{2}q_{2}^{2}}}} \nonumber \\
&& -v_{1}k_{1}\gamma_{1}\int'\frac{d\Omega}{2\pi}
\frac{d^2{\mathbf{q}}}{(2\pi)^{2}}\frac{\Omega^{2}-v_{1}^{2}
q_{1}^{2} + v_{2}^{2}q_{2}^{2}}{\left(\Omega^{2}+v_{1}^{2} q_{1}^{2}
+ v_{2}^{2}q_{2}^{2}\right)^{2}}\frac{1}{
\frac{|\mathbf{q}|}{\frac{2\pi e^{2}}{\epsilon}} + \frac{N}{8v_1
v_2}\frac{v_{1}^{2}q_{1}^{2} + v_{2}^{2}q_{2}^{2}}{\sqrt{\Omega^2 +
v_{1}^{2}q_{1}^{2} + v_{2}^{2}q_{2}^{2}}}}\nonumber \\
&&-v_{2}k_{2}\gamma_{2}\int'\frac{d\Omega}{2\pi}
\frac{d^2{\mathbf{q}}}{(2\pi)^{2}}\frac{\Omega^{2} +
v_{1}^{2}q_{1}^{2}-v_{2}^{2}q_{2}^{2}}{\left(\Omega^{2} +
v_{1}^{2}q_{1}^{2}+v_{2}^{2}q_{2}^{2}\right)^{2}}
\frac{1}{\frac{|\mathbf{q}|}{\frac{2\pi e^{2}}{\epsilon}} +
\frac{N}{8v_1 v_2}\frac{v_{1}^{2}q_{1}^{2}+v_{2}^{2}
q_{2}^{2}}{\sqrt{\Omega^2+v_{1}^{2}q_{1}^{2}+v_{2}^{2}q_{2}^{2}}}}.
\end{eqnarray}
Performing integrations according to
Eq.~(\ref{Eq:IntegrationRange}), we obtain
\begin{eqnarray}
\Sigma^{B}(\omega,\mathbf{k}) = \left(-i\omega\gamma_{0}C_{0}^{B} +
v_{1}k_{1}\gamma_{1}C_{1}^{B}+v_{2}k_{2}\gamma_{2}C_{2}^{B}\right)\ell.
\end{eqnarray}
The expressions of $C_{i}^{B}$ can be found in
Eqs.~(\ref{Eq:C0B})-(\ref{Eq:GExpressionB}).
\end{widetext}

\subsection{Contribution from disorder scattering}

The fermion self-energy generated by disorder is
\begin{eqnarray}
\Sigma_{\mathrm{dis}}(\omega) &=& \Delta
v_{\Gamma}^{2}\int'\frac{d^2 \mathbf{k}}{(2\pi)^2}\Gamma
G_{0}(\omega,\mathbf{k})\Gamma \nonumber
\\
&=&i\omega v_{\Gamma}^{2}\Delta\int'\frac{d^2\mathbf{k}}{(2\pi)^2}
\frac{\Gamma\gamma_{0}\Gamma}{\left(\omega^{2}+v_{1}^{2}k_{1}^{2} +
v_{2}^{2}k_{2}^{2}\right)}\nonumber \\
&\approx&i\omega\gamma_{0}C_{g}\ell,
\end{eqnarray}
where
\begin{eqnarray}
C_g = \frac{v_{\Gamma}^{2}\Delta}{2\pi v_{1}v_{2}}
\end{eqnarray}
for both RSP and RM, and
\begin{eqnarray}
C_g = \frac{\left(v_{\Gamma1}^{2} + v_{\Gamma2}^{2}\right)
\Delta}{2\pi v_{1}v_{2}}
\end{eqnarray}
for RVP.

\section{Corrections to fermion-disorder coupling}

The fermion-disorder coupling receives vertex corrections from three
sorts of interaction, including the Yukawa coupling, the Coulomb
interaction, and the fermion-disorder interaction, which will be
studied below.

\subsection{Vertex correction due to Yukawa coupling}

The vertex correction due to Yukawa coupling is
\begin{eqnarray}
V^{A} = -\int'\frac{d\Omega}{2\pi}\frac{d^2\mathbf{q}}{(2\pi)^2}
G_{0}(\Omega,\mathbf{q}) v_\Gamma \Gamma
G_{0}(\Omega,\mathbf{q})D^{A}(\Omega,\mathbf{q}). \nonumber \\
\end{eqnarray}
For RSP, $\Gamma=\gamma_{0}$ and we get
\begin{eqnarray}
V^{A} = v_\Gamma\gamma_{0}\left(-C_{0}^{A}\right)\ell.
\end{eqnarray}
For the two components of RVP defined by
$\Gamma=\gamma_{1}$ and $\Gamma=\gamma_{2}$, $V_{A}$ is given by
\begin{eqnarray}
V^{A} = v_\Gamma\gamma_{1}\left(-C_{1}^{A}\right)\ell,
\end{eqnarray}
and
\begin{eqnarray}
V^{A} = v_{\Gamma}\gamma_{2}\left(-C_{2}^{A}\right)\ell,
\end{eqnarray}
respectively. For RM with $\Gamma=\mathbbm{1}$, $V_{A}$ is
\begin{eqnarray}
V^{A} = v_\Gamma\mathbbm{1}\left(C_{0}^{A} + C_{1}^{A} +
C_{2}^{A}\right)\ell.
\end{eqnarray}

\subsection{Vertex correction due to Coulomb interaction}

The vertex correction due to Coulomb interaction is
\begin{eqnarray}
V^{B} &=& -\int'\frac{d\Omega}{2\pi}\frac{d^2\mathbf{q}}{(2\pi)^2}
\gamma_{0}G_{0}(\Omega,\mathbf{q}) v_\Gamma \Gamma
G_{0}(\Omega,\mathbf{q})\gamma_{0} \nonumber \\
&&\times D^{B}(\Omega,\mathbf{q}).
\end{eqnarray}
For RSP with $\Gamma=\gamma_{0}$, $V_{B}$ is
\begin{eqnarray}
V^{B} = v_\Gamma\gamma_{0}\left(-C_{0}^{B}\right)\ell.
\end{eqnarray}
For the two components of RVP defined by
$\Gamma=\gamma_{1}$ and $\Gamma=\gamma_{2}$, we obtain
\begin{eqnarray}
V^{B} = v_\Gamma\gamma_{1}\left(-C_{1}^{B}\right)\ell,
\end{eqnarray}
and
\begin{eqnarray}
V^{B} = v_{\Gamma}\gamma_{2}\left(-C_{2}^{B}\right)\ell,
\end{eqnarray}
respectively. For RM with $\Gamma=\mathbbm{1}$, we find
\begin{eqnarray}
V^{B} = v_\Gamma \mathbbm{1}\left(C_{0}^{B} - C_{1}^{B} -
C_{2}^{B}\right)\ell.
\end{eqnarray}

\subsection{Vertex correction from disorder}

The vertex correction due to disorder has the form
\begin{eqnarray}
V_{\mathrm{dis}} &=& \Delta v_{\Gamma}^{2}\int'
\frac{d^2\mathbf{p}}{(2\pi)^2} \Gamma G_0(0,\mathbf{k})
v_\Gamma\Gamma G_0(0,\mathbf{k})\Gamma\nonumber
\\
&=& v_{\Gamma}\Delta v_{\Gamma}^{2}\int
\frac{d^2\mathbf{p}}{(2\pi)^2}\frac{1}{\left(v_{1}^{2}k_{1}^{2} +
v_{2}^{2}k_{2}^{2}\right)^{2}}\nonumber \\
&& \times \Gamma\left(v_{1}k_{1} \gamma_{1} + v_{2}k_{2}
\gamma_{2}\right)\Gamma \left(v_{1}k_{1} \gamma_{1} +
v_{2}k_{2}\gamma_{2}\right)\Gamma. \nonumber \\
\end{eqnarray}
For RSP with $\gamma=\gamma_{0}$, $V_{dis}$ is
\begin{eqnarray}
V_{\mathrm{dis}} = v_{\Gamma}\gamma_{0}C_{g}\ell.
\end{eqnarray}
For the two components of RVP defined by
$\gamma=\gamma_{1}$ and $\gamma_{2}$, $V_{\mathrm{dis}}$ is
\begin{eqnarray}
V_{\mathrm{dis}} = 0.
\end{eqnarray}
For RM with $\Gamma=\mathbbm{1}$, $V_{\mathrm{dis}}$ is
\begin{eqnarray}
V_{\mathrm{dis}} = -v_{\Gamma}\mathbbm{1}C_{g}\ell.
\end{eqnarray}
\begin{widetext}

\section{Derivation of the coupled RG equations}

The action for the free fermions is given by
\begin{eqnarray}
S_{\Psi} &=& \sum_{\sigma=1}^{N}\int\frac{d\omega}{2\pi}
\frac{d^{2}\mathbf{k}}{(2\pi)^{2}} \bar{\Psi}_{\sigma}
(\omega,\mathbf{k})\left(-i\omega \gamma_{0} +
v_{1}k_{1}\gamma_{1}+v_{2}k_{2}\gamma_{2}\right)
\Psi_{\sigma}(\omega,\mathbf{k}).
\end{eqnarray}
Including the fermion self-energies induced by excitonic quantum
fluctuation, Coulomb interaction, and disorder scattering, the
action of fermions becomes
\begin{eqnarray}
S_{\Psi} &=& \sum_{\sigma=1}^{N}\int\frac{d\omega}{2\pi}
\frac{d^{2}\mathbf{k}}{(2\pi)^{2}}
\bar{\Psi}_{\sigma}(\omega,\mathbf{k}) \left[-i\omega\gamma_{0} +
v_{1}k_{1}\gamma_{1} + v_{2}k_{2}\gamma_{2} -
\Sigma^{A}(\omega,\mathbf{k})-\Sigma^{B}(\omega,\mathbf{k}) -
\Sigma_{\mathrm{dis}}(\omega)\right]\Psi_{\sigma}(\omega,\mathbf{k})\nonumber
\\
&\approx&\sum_{\sigma=1}^{N}\int\frac{d\omega}{2\pi}
\frac{d^{2}\mathbf{k}}{(2\pi)^{2}}
\bar{\Psi}_{\sigma}(\omega,\mathbf{k})\left[-i\omega\gamma_{0}
e^{\left(-C_{0}^{A}-C_{0}^{B}+C_{g}\right)\ell} + v_{1}k_{1}
\gamma_{1}e^{-\left(C_{1}^{A}+C_{1}^{B}\right)\ell} + v_{2}k_{2}
\gamma_{2}e^{-\left(C_{2}^{B}+C_{2}^{B}\right)\ell}\right]
\Psi_{\sigma}(\omega,\mathbf{k}).
\end{eqnarray}
Making the following re-scaling transformations:
\begin{eqnarray}
\omega&=&\omega'e^{-\ell}, \label{Eq:Scalingomega}
\\
k_{1}&=&k'_{1}e^{-\ell}, \label{Eq:Scalingk1}
\\
k_{2}&=&k'_{2}e^{-\ell}, \label{Eq:Scalingk2}
\\
\Psi &=& \Psi' e^{\left(2+\frac{C_{0}^{A}}{2} + \frac{C_{0}^{B}}{2}
- \frac{C_{g}}{2}\right)\ell}, \label{Eq:ScalingPsi}
\\
v_{1}&=&v_{1}'e^{\left(-C_{0}^{A}-C_{0}^{B}+C_{1}^{A} +
C_{1}^{B}+C_{g}\right)\ell}, \label{Eq:Scalingv1}
\\
v_{2}&=&v_{2}'e^{\left(-C_{0}^{A}-C_{0}^{B}+C_{2}^{A} +
C_{2}^{B}+C_{g}\right)\ell}, \label{Eq:Scalingv2}
\end{eqnarray}
the fermion action is re-written as
\begin{eqnarray}
S_{\Psi'} = \sum_{\sigma=1}^{N}\int\frac{d\omega'}{2\pi}
\frac{d^{2}\mathbf{k}'}{(2\pi)^{2}}
\bar{\Psi}_{\sigma}'(\omega',\mathbf{k}') \left[-i\omega'\gamma_{0}
+ v_{1}'k_{1}'\gamma_{1}+v_{2}'k_{2}'\gamma_{2}\right]
\Psi_{\sigma}'(\omega',\mathbf{k}'),
\end{eqnarray}
which recovers the form of the original action.

The action for the fermion-disorder coupling is
\begin{eqnarray}
S_{\mathrm{dis}} = \sum_{\sigma=1}^{N}\int\frac{d\omega}{2\pi}
\frac{d^{2}\mathbf{k}}{(2\pi)^{2}}
\int\frac{d^{2}\mathbf{k}_{1}}{(2\pi)^{2}}
\bar{\Psi}_{\sigma}(\omega,\mathbf{k})v_{\Gamma}\Gamma
\Psi_{\sigma}(\omega,\mathbf{k}_{1}) A(\mathbf{k}-\mathbf{k}_{1})
\end{eqnarray}
After taking into account the quantum corrections, it becomes
\begin{eqnarray}
S_{\mathrm{dis}} = \sum_{\sigma=1}^{N}\int\frac{d\omega}{2\pi}
\frac{d^{2}\mathbf{k}}{(2\pi)^{2}} \int\frac{d^{2}
\mathbf{k}_{1}}{(2\pi)^{2}} \bar{\Psi}_{\sigma}(\omega,\mathbf{k})
\left(v_{\Gamma}\Gamma+V^{A}+V^{B}+V_{\mathrm{dis}}\right)
\Psi_{\sigma}(\omega,\mathbf{k}_{1}) A(\mathbf{k}-\mathbf{k}_{1}).
\end{eqnarray}
In the case of RSP, we obtain
\begin{eqnarray}
S_{\mathrm{dis}} &=& \sum_{\sigma=1}^{N}\int\frac{d\omega}{2\pi}
\frac{d^{2}\mathbf{k}}{(2\pi)^{2}} \int\frac{d^{2}
\mathbf{k}_{1}}{(2\pi)^{2}} \bar{\Psi}_{\sigma}(\omega,\mathbf{k})
\left[v_{\Gamma}\gamma_{0}+v_\Gamma\gamma_{0}\left(-C_{0}^{A}\right)\ell
+v_\Gamma\gamma_{0}\left(-C_{0}^{B}\right)\ell+v_{\Gamma}\gamma_{0}C_{g}\ell\right]
\Psi_{\sigma}(\omega,\mathbf{k}_{1}) A(\mathbf{k}-\mathbf{k}_{1})
\nonumber \\
&\approx&\sum_{\sigma=1}^{N}\int\frac{d\omega}{2\pi}
\frac{d^{2}\mathbf{k}}{(2\pi)^{2}} \int\frac{d^{2}
\mathbf{k}_{1}}{(2\pi)^{2}} \bar{\Psi}_{\sigma}(\omega,\mathbf{k})
v_{\Gamma}\gamma_{0}e^{\left(-C_{0}^{A}-C_{0}^{B}+C_{g}\right)\ell}
\Psi_{\sigma}(\omega,\mathbf{k}_{1}) A(\mathbf{k}-\mathbf{k}_{1}).
\label{Eq:SdisCorrectRCP}
\end{eqnarray}
For the two components of RVP, $S_{\mathrm{dis}}$ is expressed as
\begin{eqnarray}
S_{\mathrm{dis}} &=& \sum_{\sigma=1}^{N}\int\frac{d\omega}{2\pi}
\frac{d^{2}\mathbf{k}}{(2\pi)^{2}} \int\frac{d^{2}
\mathbf{k}_{1}}{(2\pi)^{2}}
\bar{\Psi}_{\sigma}(\omega,\mathbf{k})\left[v_{\Gamma}\gamma_{1}
+v_\Gamma\gamma_{1}\left(-C_{1}^{A}\right)\ell+v_\Gamma\gamma_{1}
\left(-C_{1}^{B}\right)\ell\right]
\Psi_{\sigma}(\omega,\mathbf{k}_{1}) A(\mathbf{k}-\mathbf{k}_{1})
\nonumber \\
&\approx&\sum_{\sigma=1}^{N}\int\frac{d\omega}{2\pi}
\frac{d^{2}\mathbf{k}}{(2\pi)^{2}}
\int\frac{d^{2}\mathbf{k}_{1}}{(2\pi)^{2}}
\bar{\Psi}_{\sigma}(\omega,\mathbf{k})v_{\Gamma}\gamma_{1}
e^{-\left(C_{1}^{A}+C_{1}^{B}\right)\ell}
\Psi_{\sigma}(\omega,\mathbf{k}_{1})A(\mathbf{k}-\mathbf{k}_{1}),
\label{Eq:SdisCorrectRGP1}
\end{eqnarray}
and
\begin{eqnarray}
S_{\mathrm{dis}} &=& \sum_{\sigma=1}^{N}\int
\frac{d\omega}{2\pi}\frac{d^{2}\mathbf{k}}{(2\pi)^{2}}
\int\frac{d^{2}\mathbf{k}_{1}}{(2\pi)^{2}}
\bar{\Psi}_{\sigma}(\omega,\mathbf{k})\left[v_{\Gamma}\gamma_{2} +
v_\Gamma\gamma_{2}\left(-C_{2}^{A}\right)\ell +
v_\Gamma\gamma_{2}\left(-C_{2}^{B}\right)\ell\right]
\Psi_{\sigma}(\omega,\mathbf{k}_{1}) A(\mathbf{k}-\mathbf{k}_{1})
\nonumber \\
&\approx&\sum_{\sigma=1}^{N}\int\frac{d\omega}{2\pi}
\frac{d^{2}\mathbf{k}}{(2\pi)^{2}}
\int\frac{d^{2}\mathbf{k}_{1}}{(2\pi)^{2}}
\bar{\Psi}_{\sigma}(\omega,\mathbf{k})v_{\Gamma}\gamma_{2}
e^{-\left(C_{2}^{A}+C_{2}^{B}\right)\ell}
\Psi_{\sigma}(\omega,\mathbf{k}_{1})A(\mathbf{k}-\mathbf{k}_{1}),
\label{Eq:SdisCorrectRGP2}
\end{eqnarray}
respectively. For RM, $S_{\mathrm{dis}}$ is cast in the form
\begin{eqnarray}
S_{\mathrm{dis}}&=&\sum_{\sigma=1}^{N}\int\frac{d\omega}{2\pi}
\frac{d^{2}\mathbf{k}}{(2\pi)^{2}}
\int\frac{d^{2}\mathbf{k}_{1}}{(2\pi)^{2}}
\bar{\Psi}_{\sigma}(\omega,\mathbf{k})\left[v_{\Gamma}
\mathbbm{1}+v_\Gamma\mathbbm{1}\left(C_{0}^{A}+C_{1}^{A} +
C_{2}^{A}\right)\ell + v_\Gamma\mathbbm{1} \left(C_{0}^{B} -
C_{1}^{B}-C_{2}^{B}\right)\ell-v_{\Gamma}\mathbbm{1}C_{g}\ell
\right]\nonumber \\
&&\times\Psi_{\sigma}(\omega,\mathbf{k}_{1})
A(\mathbf{k}-\mathbf{k}_{1})\nonumber \\
&=&\sum_{\sigma=1}^{N}\int\frac{d\omega}{2\pi}
\frac{d^{2}\mathbf{k}}{(2\pi)^{2}} \int\frac{d^{2}
\mathbf{k}_{1}}{(2\pi)^{2}} \bar{\Psi}_{\sigma}(\omega,\mathbf{k})
v_{\Gamma}\mathbbm{1}e^{\left(C_{0}^{A}+C_{1}^{A}+C_{2}^{A}+C_{0}^{B}
- C_{1}^{B}-C_{2}^{B} -C_{g}\right)\ell}
\Psi_{\sigma}(\omega,\mathbf{k}_{1})A(\mathbf{k}-\mathbf{k}_{1}).
\label{Eq:SdisCorrectRM}
\end{eqnarray}
We then employ the re-scaling transformations given by
Eqs.~(\ref{Eq:Scalingomega})-(\ref{Eq:ScalingPsi}). The random
potential $A(\mathbf{k})$ should be re-scaled as follows
\begin{eqnarray}
A(\mathbf{k}) = A'(\mathbf{k'})e^{\ell}.
\end{eqnarray}
The parameter $v_{\Gamma}$ is re-scaled as
\begin{eqnarray}
v_{\Gamma}=v_{\Gamma}' \label{Eq:ScalingvGammaRCP}
\end{eqnarray}
for Eq.~(\ref{Eq:SdisCorrectRCP}),
\begin{eqnarray}
v_{\Gamma}=v_{\Gamma}'e^{\left(-C_{0}^{A}-C_{0}^{B}+C_{1}^{A} +
C_{1}^{B}+C_{g}\right)\ell} \label{Eq:ScalingvGammaRVP1}
\end{eqnarray}
for Eq.~(\ref{Eq:SdisCorrectRGP1}),
\begin{eqnarray}
v_{\Gamma}=v_{\Gamma}'e^{\left(-C_{0}^{A}-C_{0}^{B}+C_{2}^{A} +
C_{2}^{B}+C_{g}\right)\ell} \label{Eq:ScalingvGammaRVP2}
\end{eqnarray}
for Eq.~(\ref{Eq:SdisCorrectRGP2}), and
\begin{eqnarray}
v_{\Gamma}=v_{\Gamma}'e^{\left(-2C_{0}^{A}-C_{1}^{A}-C_{2}^{A} -
2C_{0}^{B}+C_{1}^{B}+C_{2}^{B} +2C_{g}\right)\ell}
\label{Eq:ScalingvGammaRM}
\end{eqnarray}
for Eq.~(\ref{Eq:SdisCorrectRM}). After carrying out the above
manipulations, we re-write the action for fermion-disorder coupling
as follows
\begin{eqnarray}
S_{\mathrm{dis}} = \sum_{\sigma=1}^{N}\int\frac{d\omega'}{2\pi}
\frac{d^{2}\mathbf{k}'}{(2\pi)^{2}} \int\frac{d^{2}
\mathbf{k}_{1}'}{(2\pi)^{2}}
\bar{\Psi}_{\sigma}'(\omega',\mathbf{k}')v_{\Gamma}'\mathbbm{1}
\Psi_{\sigma}'(\omega',\mathbf{k}_{1}')A'(\mathbf{k}'-\mathbf{k}_{1}'),
\end{eqnarray}
which restores the form of the original action.

From Eqs.~(\ref{Eq:ScalingPsi}), we obtain the RG equation for the
quasiparticle $Z_{f}$
\begin{eqnarray}
\frac{dZ_{f}}{d\ell} = \left(C_{0}^{A} + C_{0}^{B} -
C_{g}\right)Z_{f}.
\end{eqnarray}
According to Eqs.~(\ref{Eq:Scalingv1}) and (\ref{Eq:Scalingv2}), the
RG equations for $v_{1}$ and $v_{2}$ are given by
\begin{eqnarray}
\frac{dv_{1}}{d\ell} &=& \left(C_{0}^{A}+C_{0}^{B}-C_{1}^{A} -
C_{1}^{B}-C_{g}\right)v_{1},
\\
\frac{dv_{2}}{d\ell} &=& \left(C_{0}^{A}+C_{0}^{B}-C_{2}^{A} -
C_{2}^{B}-C_{g}\right)v_{2}.
\end{eqnarray}
The RG equation for the velocity ratio $v_{2}/v_{1}$ can be readily
derived:
\begin{eqnarray}
\frac{d\left(v_{2}/v_{1}\right)}{d\ell} &=&
\frac{\frac{dv_{2}}{dl}v_{1}-v_{2}\frac{dv_{1}}{dl}}{v_{1}^{2}}
\nonumber \\
&=& \left(C_{1}^{A}-C_{2}^{A}+C_{1}^{B}-C_{2}^{B}\right)
\frac{v_{2}}{v_{1}}.
\end{eqnarray}
Based on
Eqs.~(\ref{Eq:ScalingvGammaRCP})-(\ref{Eq:ScalingvGammaRM}), we
obtain the RG equation for the parameter $v_{\Gamma}$
\begin{eqnarray}
\left\{\begin{array}{ll} \frac{d v_{\Gamma}}{d\ell}=0 &\texttt{RSP},
\\
\\
\frac{dv_{\Gamma}}{d\ell}=\left(C_{0}^{A}+C_{0}^{B}-C_{1}^{A} -
C_{1}^{B}-C_{g}\right)v_{\Gamma} & \gamma_{1} \ \texttt{component of
RVP}
\\
\\
\frac{dv_{\Gamma}}{d\ell}=\left(C_{0}^{A}+C_{0}^{B}-C_{2}^{A} -
C_{2}^{B}-C_{g}\right)v_{\Gamma} & \gamma_{2} \ \texttt{component of
RVP}
\\
\\
\frac{dv_{\Gamma}}{d\ell}=\left(2C_{0}^{A}+C_{1}^{A}+C_{2}^{A} +
2C_{0}^{B}-C_{1}^{B}-C_{2}^{B} -2C_{g}\right)v_{\Gamma} &\texttt{RM}
\end{array}\right.
\end{eqnarray}
\end{widetext}

\end{document}